\documentclass[11pt,a4paper]{article}
\usepackage{jheppub}
\usepackage{comment}

\def\be{\begin{equation}}
\def\ee{\end{equation}}
\def\ba{\begin{eqnarray}}
\def\ea{\end{eqnarray}}
\def\bi{\begin{itemize}}
\def\ei{\end{itemize}}
\def\nn{\nonumber}

\def\xh{\hat{x}}
\def\qh{\hat{q}}

\def\wb{\bar{w}}
\def\zb{\bar{z}}
\def\w{\omega}
\def\t{\tau}

\def\D{\mathcal{D}}

\def\G{\mathcal{G}}

\def\e{\epsilon}
\def\Qs{Q^{\text{soft}}}
\def\Qh{Q^{\text{hard}}}

\def\M{\mathcal{M}}
\def\phifree{\varphi_{\text{free}}}
\def\free{\text{free}}
\def\inn{\text{in}}
\def\out{\text{out}}
\def\bra{\langle}
\def\ket{\rangle}
\def\hom{{\rm hom}}
\def\scri{\mathcal{I}}
\def\kf{\text{KF}}
\def\A{\mathcal{A}}
\def\soft{\text{soft}}

\title{Loop Corrected Soft Photon Theorem as a Ward Identity}

\author[a]{Miguel Campiglia} 
\author[b]{Alok Laddha} 
\affiliation[a]{Instituto de F\'isica, Facultad de Ciencias,  Montevideo 11400, Uruguay}
\affiliation[b]{Chennai Mathematical Institute, Siruseri, Chennai, India}

\emailAdd{campi@fisica.edu.uy}
\emailAdd{aladdha@cmi.ac.in}

\abstract{
Recently Sahoo and Sen obtained a series of remarkable results concerning  sub-leading soft photon and graviton theorems in four dimensions. Even though the S-matrix is infrared divergent, they have shown that the subleading soft theorems are well defined and exact statements in QED and perturbative Quantum  Gravity. However unlike the well studied Cachazo-Strominger soft theorems in  tree-level amplitudes, the new subleading soft expansion is at the order $\ln\omega$ (where $\omega$ is the soft frequency)  and the corresponding soft factors structurally show completely different properties then their tree-level counterparts. Whence it is natural to ask if these theorems are associated to asymptotic symmetries of the S-matrix.

We consider this question in the context of sub-leading soft photon theorem  in scalar QED and show that there are indeed an infinity of conservation laws whose Ward identities are equivalent to the loop-corrected soft  photon theorem. This shows that in the case of four dimensional QED, the leading and sub-leading soft photon theorems are equivalent to Ward identities of (asymptotic) charges.
}

\begin{document}

\maketitle


\tableofcontents

\section{Introduction}

In last few years there has been remarkable progress in our understanding of the symmetry structure of gauge theories and  gravity. This understanding is based on Strominger's key insight that soft theorems in these theories are manifestations of Ward identities of  so-called asymptotic symmetries. For theorems like Weinberg soft graviton theorem, this relationship implies that BMS group is an exact symmetry of the (perturbative) Gravitational S-matrix in four dimensions. However in four spacetime dimensions, the issue is more subtle when it comes to sub-leading soft theorems. This is because sub-leading soft theorems in QED or in gravity are sensitive to the infrared structure of the S-matrix unlike the leading soft theorems.
In fact in \cite{bern-loop} it was shown that sub-leading soft theorems in Gauge theories and Gravity  are infrared divergent. In this analysis it was assumed that the soft expansion was  a power series in $\omega$ and infrared modifications were then derived for the soft factors in dimensional regularisation. In \cite{stromloop} Such modifications to the soft factors  were beautifully understood in terms of Ward identities. That is, it was shown that the same asymptotic symmetries whose charges gave rise to tree-level soft theorems gave rise to infrared modified soft theorems due to change in the Asymptotic charges at loop level. 

In the seminal work \cite{sahoo}, Sahoo and Sen  have shown that if we do not insist on the power series expansion in soft frequency, then, in four spacetime dimensions the subleading soft theorems in QED and gravity have a markedly different structure  to the corresponding theorems in tree-level amplitudes. More in detail, it is by now well understood that tree level S-matrix has a soft expansion which is a Laurent series in the soft frequency $\omega$. In fact, understanding of such tree-level soft theorems  as Ward identities associated to (Spontaneously broken) Asymptotic symmetries have been investigated in a large number of recent works (see \cite{strominger} and \cite{ash-campi} for recent reviews). Sahoo and Sen have shown that once loop corrections are taken into account, the situation changes rather drastically and the soft expansion goes as $\frac{1}{\omega}S^{(0)}\ +\ \ln\omega\ S^{(1)}\ +\ \dots$. In particular, in scalar QED, the soft theorem is given by

\be\label{ss-log}
{\cal M}_{n+1}(p_{1},\dots, p_{n},k)\ =\ \left[\frac{1}{\omega}\ S^{(0)}\ -\ \ln\omega\ S^{(\ln)}\ \right]{\cal M}_{n}(p_{1},\dots, p_{n})\ +\ \dots
\ee

where $S^{(0)}$ is the Weinberg soft factor and the loop-corrected soft factor $S^{(\ln)}$ is given by 

\be\label{S-ln}
S^{(\ln)}\ =\ \sum_{a,b} f_{1}(p_{a},p_{b})\ +\ f_{2}(p_{a},p_{b})
\ee
with
\be
\begin{array}{lll}
f_{1}(p_{a},p_{b})\ &&=\ \frac{-i}{4\pi}q_{a}^{2}q_{b}\ m_{a}^{2}m_{b}^{2}\ \frac{\epsilon_{\mu}\ (p_{a}^{\mu}p_{b}^{\rho}\ -\ p_{a}^{\rho}p_{b}^{\mu}) k_{\rho}}{[(p_{a}\cdot p_{b})^{2}\ -\ m_{a}^{2}m_{b}^{2}]^{\frac{3}{2}}}\ \textrm{if both (a,b) are in-coming or outgoing}\\
\vspace*{0.1in}
&& =\ 0\ \textrm{otherwise},\\
\vspace*{0.2in}
f_{2}(p_{a},p_{b})\ &&=\ \left[-\frac{q_{a}^{2}q_{b}}{8\pi^{2}}\ \ln[\frac{p_{a}\cdot p_{b}\ +\ \sqrt{(p_{a}\cdot p_{b})^{2}\ -\ m_{a}^{2}m_{b}^{2}}}{p_{a}\cdot p_{b}\ +\ \sqrt{(p_{a}\cdot p_{b})^{2}\ -\ m_{a}^{2}m_{b}^{2}}}]\ \frac{m_{a}^{2}m_{b}^{2}}{\{(p_{a}\cdot p_{b})^{2}\ -\ m_{a}^{2}m_{b}^{2}\}^{\frac{3}{2}}}\ \left\{-\epsilon\cdot p_{b}\ +\ \frac{\epsilon\cdot p_{a}}{k\cdot p_{a}}\ k\cdot p_{b}\right\}\right.\\
\vspace*{0.1in}
&&\left.+\ \frac{q_{a}^{2}q_{b}}{4\pi^{2}}\ \frac{p_{a}\cdot p_{b}}{(\ (p_{a}\cdot p_{b})^{2}\ -\ m_{a}^{2}m_{b}^{2}\ )}\ +\ \left\{-\epsilon\cdot p_{b}\ +\ \frac{\epsilon\cdot p_{a}}{k\cdot p_{a}}\ k\cdot p_{b}\right\}\right]\\
\vspace*{0.1in}
&&\hspace*{1.9in} \textrm{for all pairs of particles, (a,b)}
\end{array}
\ee
where $\epsilon^{\mu}$ is the polarization of the soft photon with momentum $k$. In \cite{sahoo}, the authors referred to $\sum_{a,b}\ f_{1}(p_{a},p_{b})$ as  classical soft factor since the $\ln\omega$ term in the soft expansion of classical  electromagnetic radiation is dictated by this factor. 

If the infinite dimensional asymptotic symmetries are symmetries of perturbative QED as opposed to tree-level S matrix, then it must be so that the logarithmic sub-leading soft photon theorem is equivalent to the corresponding Ward identities. In this paper, we take a first step towards understanding such symmetries in the Quantum (as opposed to tree-level) regime. Namely, we show that even in perturbative QED beyond tree-level there exist an infinite number of charges (that we refer to as sub-leading charges) which are conserved in the sense that the charge at  the asymptotic past  equals the charge at the asymptotic future. Even though the structure of log-corrected sub-leading soft photon theorem is rather involved, rather remarkably it turns out the Ward identities associated to quantized sub-leading charges are equivalent to the sub-leding soft photon theorem. This is our main result. We also show that the our entire formalism (although not aimed to cure infra-red divergences of the S matrix) is such that the sub-leading charges have a rather direct relationship with Kulish-Faddeev dressing of charged states.


The plan of the paper is as follows. We first show that the subleading soft photon theorem in tree-level scattering amplitudes in scalar QED is equivalent to Ward identities associated to symmetries which are parametrized by vector fields on $S^{2}$. 
We then show that the asymptotic charges associated to these symmetries get corrected at loop level and derive the charges. As is always the case, this charge is a sum of a so-called soft charge  and a hard charge which depends on the three point coupling of matter with the asymptotic gauge field. Unlike the tree-level soft charge, the soft charge in the present case is such that it isolates the $\ln\omega$ frequency mode in the soft expansion.  We show how the Ward identities associated to these charges are equivalent to the subleading soft photon theorem derived by Sahoo and Sen. As these theorems are 1-loop exact, our analysis reveals that there are infinity of conserved charges in perturbative scalar QED, one of which generate  large $U(1)$ gauge transformation and the other  parametrized by vector fields on the celestial sphere.

\section{Anatomy of fall-offs and Asymptotic charges}

The essential steps in which we are led to different Asymptotic conditions for gauge fields and matter currents are summarized below. Our summary here is brief and necessarily qualitative as we would like to give a coherent picture of the Asymptotics at time-like and Null infinity. Details of the structure presented below are in the main body of the paper. 

\begin{itemize}
\item We start with the basic observation that Coulombic modes of scalar potential $A_{\tau}$ decay as $\frac{1}{\tau}$ at late times. (Here $\tau$ parametrizes  asymptotic proper time of time-like geodesics.)
\item Due to this slowly decaying Coulombic mode, Asymptotic behaviour of scalar field at late times is modified such that the spatial matter current components decay as $\frac{\ln\tau}{\tau^{3}}$. 
\item This logarithmic decay of current components implies that the Maxwell fields (e.g. Electric field) decays as $\frac{1}{\tau} \vec{A}(y)\ +\ \frac{\ln\tau}{\tau^{2}}\ \vec{B}(y)$ at large $\tau$ with $y$ being the co-ordinates on fixed time slice. 
\item The behaviour of Maxwell fields near time-like infinity can be used to determine the fall-off conditions of these fields in the future of Null infinity. In particular, the above mentioned late time decay of Maxwell fields, along with Maxwell's equations can be used to determine $\vert u\vert\rightarrow\infty$ behaviour of the radiative data $A_{A}(u,\hat{x})$ at Null infinity  and is given by $\overset{0}{A}_{A}\ +\ \frac{1}{\vert u\vert}\overset{1}{A}_{A}$. 
\item The above fall-off of radiative data at large $u$ implies that in addition to infinity of leading charges, one has an infinity of non-trivial ``sub-leading" charges which are defined at $u\ =\ -\infty$ in terms of $\overset{1}{A}_{A}(\hat{x})$.
\item We then show that similar sub-leading charges are also non-trivial at $v\ =\ +\infty$ on the boundary of past Null infinity and in fact due to Equations of motion of the theory at spatial infinity imply that these two sets of charges are equal, leading to new set of conservation laws. 
\end{itemize}

\section{Asymptotics of classical electrodynamics at time and null infinities} \label{treesec}
\subsection{Fall-offs at time-infinity}
Massive charged  fields decay very rapidly ($O(\frac{1}{r^{\infty}})$) at null infinity \cite{winicour-1988} and hence their asymptotic modes are best described by analysing their behaviour at time-like infinity. Details of this construction were given in \cite{massive} and here we summarise the key aspects of the construction.

In the usual conformal compactification scheme of Penrose, future and past time-like infinities $i^{\pm}$ are points and as such they are ill suited to understand late time decay of massive fields. A better arena for this purpose is ``a blow up of $i^{\pm}$" which is a three dimensional space-like hyperboloid.

This hyperboloid picture is manifest in the following coordinates adapted to time infinity,
\be
\t:= \sqrt{t^2-r^2}, \quad \rho := \frac{r}{\sqrt{t^2-r^2}}
\ee
in terms of which the Minkowski line element takes the form
\be
ds^2= -d \t^2 + \t^2 h_{\alpha \beta} d y^\alpha d y^\beta
\ee
where $y^\alpha=(\rho,x^A)$ are coordinates on the `time-infinity' hyperboloid with metric
\be
h_{\alpha \beta} d y^\alpha d y^\beta = \frac{d \rho^2}{1+\rho^2}+\rho^2 \gamma_{AB} dx^A dx^B ,
\ee
 and $\gamma_{AB}$ the unit-sphere metric. We will denote by
\be
Y^\mu= (\sqrt{1+\rho^2}, \rho \xh)
\ee
the unit vector on Minkowski space parametrized by $y^\alpha$. A spacetime point $x^\mu$ in hyperbolic coordinates is then written as
\be
x^\mu = \t Y^\mu
\ee
For concreteness we will focus on future time infinity but similar considerations apply to past time infinity.

In Lorenz gauge, the fall-offs of Maxwell fields (sourced by massive charged matter) at time-like infinity are given by\footnote{The $1/\t$ fall-off of the scalar potential ${\cal A}_{\tau}$ is a Coulomb effect and can be deduced from the fact that when sourced by massive particles with asymptotic constant velocities, the $\frac{1}{r}$ potential decays as $1/t \sim 1/\t$. See e.g. \cite{KF}.}
\begin{equation} \label{fallAtime}
\begin{array}{lll}
{\cal A}_{\tau}\ =\ \frac{1}{\tau}A_\t\ +\  \cdots \\
\vspace*{0.1in}
{\cal A}_{\alpha}\ =\ A_{\alpha}\ +\ \cdots,
\end{array}
\end{equation}
where the dots denotes terms that are subleading for  $\t \to \infty$. In \cite{massive} it was shown that these  fall-offs 
imply that the $\t \to \infty$ asymptotics  of the scalar field $\varphi$ has a $\ln \t$ phase relative to the free field expression, i.e. the solution to the field equations of a charged scalar field coupled to the Maxwell field has the following fall-off at future time-like infinity $i^{+}$, 
\be
\varphi(\t,y) \stackrel{\t \to \infty}{=} e^{i e \ln \t  A_\t(y)} \phifree(\t,y) \label{dressedphi}
\ee
where the free field admits the expansion 
\be
\phifree(\t,y) = \frac{\sqrt{m}}{2 (2 \pi \tau)^{3/2}}  \left( b(y) e^{ -i \t m } +  c^*(y) e^{ i \t m } \right) + \cdots .\label{freephi}
\ee
As we shall discuss in section \ref{KFsec},  in quantum theory the asymptotics (\ref{dressedphi}) corresponds to the Kulish-Faddeev asymptotics of quantum charged fields \cite{KF}. As is common in the literature, we call the logarithmic phase the ``dressing'', and refer to the field (\ref{dressedphi}) as the ``dressed field''.

The asymptotic form of the dressed field implies the following $\t \to \infty$ fall-offs for the matter current, 
\be
\begin{array}{lll} \label{fallJtime}
j_\t(\t,y) = \frac{1}{\t^3}\overset{0}{j}_\t(y) + \frac{\ln \t}{\t^4} \overset{\ln}{j}_\t(y) + \frac{1}{\t^{4}}\overset{1}{j}_\t(y)\\
j_\alpha(\t,y) =  \frac{\ln \t}{\t^3}\overset{\ln}{j}_\alpha(y) + \frac{1}{\t^3}\overset{0}{j}_\alpha(y) + \ldots.
\end{array}
\ee
These in turn  dictate the asymptotic fall-offs of $F_{ab}$ at time-like infinity. The  $\tau \to \infty$ asymptotic expansion compatible with (\ref{fallAtime}) and (\ref{fallJtime})  is found to be given by

\ba\label{jan28-1}
F_{ \t \alpha}(\t,y) & =& \frac{1}{\t} \overset{0}{F}_{\t \alpha }(y) + \frac{\ln \t}{\t^2}\overset{\ln}{F}_{\t \alpha }(y) + \ldots \\
F_{\alpha \beta}(\t,y) & =& \overset{0}{F}_{\alpha \beta}(y) + \frac{\ln \t}{\t} \overset{\ln}{F}_{\alpha \beta}(y) + \ldots.
\ea
We can now solve Maxwell equations at each order in the $\frac{1}{\tau}$ and $\frac{\ln\tau}{\tau}$ expansion. It is straightforward to check that for the leading components $\overset{0}{F}_{ab}$ we get equations which are independent of the $\ln\tau$ terms in the current,

\begin{equation}
\begin{array}{lll}
{\cal D}^{\alpha}\overset{0}{F}_{\tau\alpha}\ =\ \overset{0}{j}_{\tau}\nonumber\\
{\cal D}^{\beta}\overset{0}{F}_{\alpha\beta}\ =\ 0,
\end{array}
\end{equation}
where $\D_\alpha$ is the covariant derivative on the hyperboloid.
For the $\overset{\ln}{F}_{ab}$ terms Maxwell equations take the form
\be
\partial_{[\alpha} \overset{\ln}{F}_{\beta \gamma]}=0, \quad \partial_\alpha \overset{\ln}{F}_{\t \beta} - \partial_\beta \overset{\ln}{F}_{\t \alpha}+ \overset{\ln}{F}_{\alpha \beta}=0, \quad \D^\alpha \overset{\ln}{F}_{\t \alpha} = \overset{\ln}{j}_\t , \quad \D^\beta \overset{\ln}{F}_{\alpha \beta} -  \overset{\ln}{F}_{\t \alpha} = \overset{\ln}{j}_\alpha \label{ecln}
\ee
whereas  current conservation $\nabla^a j_a=0$ implies  
\be \label{divjln}
\overset{\ln}{j}_\t(y) + \D^\alpha \overset{\ln}{j}_\alpha(y) =0.
\ee
 Using the above equations it can be shown that $\overset{\ln}{F}_{\tau\alpha}$ is determined in terms of the $\ln$-currents as
\begin{equation}\label{feb1-3}
\left[{\cal D}^{2}\ +\ 1\right]\overset{\ln}{F}_{\tau\alpha}\ =\ \overset{\ln}{j}_{\alpha}\ +\ {\cal D}_{\alpha}\overset{\ln}{j}_{\tau} .   
\end{equation}

We now note that all the terms proportional to $\ln\tau$ in the matter current arise due to the $\ln\tau$ term of the asymptotic matter field. In an expansion in the coupling constant $e$, these terms are necessarily higher order as compared to the leading order current.  Whence if we are also interested in the expansion in the coupling constant, then at leading order in the coupling we can ignore the log terms and all the components of the currents admit a power-law fall off. 
As we will see in  section \ref{sec4}, these ``$O(e)$ fall-offs''
lead  to the conserved asymptotic charges associated with Low's soft photon theorem in tree-level amplitudes. 

\subsection{$O(e)$ fall-offs at time and null infinities}
At $O(e)$ in the coupling constant, the current can be computed from the free scalar field 
with no 
dressing accompanying it. In this case we have
\ba
j_\t(\t,y) & = & \frac{1}{\t^3} \overset{0}{j}_\t(y) + \frac{1}{\t^4} \overset{1}{j}_\t(y) + \cdots  \\
j_\alpha(\t,y)  &= & \frac{1}{\t^3}\overset{0}{j}_\alpha(y) + \frac{1}{\t^4}\overset{1}{j}_\alpha(y) + \cdots
\ea
where as before, $y$ denotes a point on the time-infinity hyperboloid. 

Such power-law fall-offs for the current imply  the field $F_{ab}$ also falls off as a Laurent series in $\frac{1}{\tau}$ obtained by setting all the log terms in Eq.(\ref{jan28-1}) to zero:

\ba\label{jan28-2}
F_{ \t \alpha}(\t,y) & =& \frac{1}{\t} \overset{0}{F}_{\t \alpha }(y) + \frac{1}{\t^2}\overset{1}{F}_{\t \alpha }(y) + \cdots \\
F_{\alpha \beta}(\t,y) & =& \overset{0}{F}_{\alpha \beta}(y) + \frac{1}{\t} \overset{1}{F}_{\alpha \beta}(y) + \cdots.
\ea

Maxwell's equations and Bianchi identities once again determine $\overset{n}{F}_{ab}$ in terms of matter currents $\overset{n}{j}_{a}$. In particular, the subleading field $\overset{1}{F}_{\tau\alpha}$ 
satisfies \cite{subn}
\begin{equation} \label{poisson}
\left[{\cal D}^{2}\ +\ 1\right]\overset{1}{F}_{\tau\alpha}\ =\ \overset{0}{j}_{\alpha}\ +\ {\cal D}_{\alpha}\overset{1}{j}_{\tau},
\end{equation}
where
\be
\overset{1}{j}_\t= -  \D^\alpha \overset{0}{j}_\alpha  \label{jtaudiv}
\ee
from current conservation. 
These equations imply that the  $\rho \to \infty$ behavior of the $\overset{1}{F}_{ab}$ fields is given by (see Appendix \ref{greenapp})
\ba
\overset{1}{F}_{\t A}(\rho,\xh) = \rho^{-1} \overset{1,0}{F}_{\t A}(\xh)+ \cdots \label{falltauA}\\
\overset{1}{F}_{\t \rho}(\rho,\xh) = \rho^{-4} \overset{1,0}{F}_{\t \rho}(\xh)+ \cdots \label{falltaurho}
\ea
with
\be
- \overset{1,0}{F}_{\t \rho} + D^A \overset{1,0}{F}_{\t A} =0,
\ee
where we assumed  that $\overset{1}{j}_\t(\rho,\xh)$ and $\overset{1}{j}_\alpha(\rho,\xh)$ vanish sufficiently fast for $\rho \to \infty$ so that  we can use the free field equations in this limit. 

We now discuss what these fall-offs imply at null-infinity. As we shall see, the asymptotic behaviour of the fields from the perspective of time-like infinity dovetails perfectly with the ``tree-level" fall-offs of radiative fields at future null infinity ${\cal I}^{+}$ discussed in \cite{subn}.  

Recall that near future null-infinity, Maxwell fields are expanded in inverse powers of $r$, with the coefficients of the expansion being functions of retarded time $u=t-r$. For instance, 
\be
{\cal F}_{rA}(r,u,\xh)  =   \frac{1}{r^2} \overset{0}{F}_{rA}(u,\xh) + O(1/r^3).
\ee
Maxwell equations can then be solved recursively in terms of the `free data' $A_A(u,\xh)$, the transverse gauge potential at ${\cal I}^+$. 

In \cite{subn} it was shown that corresponding to the tree-level soft expansion of the photon field, the free data at ${\cal I}^{+}$ has  $u \to \pm \infty$ fall-offs such that $\partial_u A_A(u,\xh)$ decays faster than any power of $1/u$,
\be
A_A(u,\xh) \stackrel{|u| \to \infty}{=} O(1) + O(1/|u|^\infty). \label{Oefallu}
\ee
This in turn determines the $u \to \infty$ fall-offs for  $\overset{0}{F}_{ab}(u,\xh)$, $\overset{1}{F}_{ab}(u,\xh)$, etc. For instance, using the leading Maxwell equations $\nabla^b F_{ub}=0$,   $\nabla^b F_{Ab}=0$:
\be
\begin{array}{r}
\partial_u \overset{0}{F}_{ru} + \partial_u D^A A_A =0 \\
\partial_u  \overset{0}{F}_{rA} + \frac{1}{2} D_A \overset{0}{F}_{ru} -   D^B D_{[A}A_{B]} =0
\end{array} \label{leadingmaxwell}
\ee
one can easily show that (\ref{Oefallu}) implies  
 \be 
\overset{0}{F}_{r A}(u,\xh)  \stackrel{u \to  \pm\infty}{=}  u \overset{0,0}{F^\pm}_{rA}(\xh)  + \overset{0,1}{F^\pm}_{rA}(\xh) + O(1/u^\infty)  \label{OefalluFrA}
\ee
where the coefficients appear as `integration constants' when solving the field equations.

Now,  the celestial sphere on ${\cal I}^{+}$ at $u=\infty$ can be thought of as the  $\rho\rightarrow\infty$ boundary of  the asymptotic hyperboloid at $\tau\ =\ \infty$. In particular, given the asymptotic expansion at large $\tau$ we can derive corresponding  $u \to \infty$ fall-offs at null infinity. Using an analysis exactly analogous to the one in \cite{subn}\footnote{There the analysis related fields at the (future) boundary of spatial infinity with with fields at $u=-\infty$.} it can be shown that 
the $O(e)$ fall-offs given in (\ref{jan28-2}) imply the fall-offs (\ref{OefalluFrA}) at null infinity. Furthermore, the `integration constants' in (\ref{OefalluFrA})  are determined in terms of the asymptotic fields at time-like infinity. In particular one finds
\be
   \overset{0,1}{F^+}_{r A}(\xh) = \overset{1,0}{F}_{\t A}(\xh) , \label{bdytiA}
\ee
where $\overset{1,0}{F}_{\t A}(\xh)$ is the leading $\rho \to \infty$ coefficient of $\overset{1}{F}_{\t A}(y)$ as  defined in Eq. (\ref{falltauA}).

\subsection{Fall-offs at null infinity}

As we saw in the previous subsection, ignoring logarithmic terms (which were higher order in the coupling constant) in the matter fields was tantamount to considering the $O(e)$ expansion of radiative fields at null infinity. We  now consider the full $\t \to \infty$ asymptotics including logarithmic terms, and study its implications at null-infinity.  

We again focus on $F_{rA}$.  Starting from the fall-offs at time-infinity (\ref{jan28-1}), and changing coordinates from $(\t,\rho)$ to $(r,u)$ one finds 
\be\label{jan28-5}
{\cal F}_{rA} = \frac{\ln r}{r^2}  \overset{\ln}{F}_{r A} +\frac{1}{r^2} \overset{0}{F}_{r A}  + \cdots
\ee
with $\overset{\ln}{F}_{r A} \stackrel{u \to \infty}{=} O(u^0)$ and
\be\label{jan28-6}
\overset{0}{F}_{r A}(u,\xh) \stackrel{u \to  \infty}{=}  u \overset{0,0}{F^+}_{rA}(\xh) + \ln u \overset{0,\ln}{F^+}_{rA}(\xh) + O(u^{0}) . 
\ee
There are thus two modifications from the previous asymptotics: (1) a  $\ln r/r^2$ term in ${\cal F}_{rA}$ and (2) a $\ln u$ term in $\overset{0}{F}_{r A}$. The first modification does not play much of a role: The field equations at null infinity imply this term is $u$-independent,
\be
\partial_u \overset{\ln}{F}_{r A} =0,
\ee
The second modification can be understood as a change in the $u \to \infty$ fall-offs of  the free data $A_A(u,\xh)$, from (\ref{Oefallu}) to
\be\label{feb2-1}
A_A(u,\xh) \stackrel{u \to \infty}{=} O(1) + O(1/u) + \cdots.
\ee
The appearance of a $O(1/u)$ `tail' is in agreement with the expected asymptotics of radiated fields due to classical scattering particles \cite{observational}, which in turn corresponds to a $\ln \omega$ subleading term in the low frequency expansion \cite{sahoo}. Thus, the null-infinity asymptotics ($u\ \rightarrow\ \infty$) that we find may be thought of as giving ``loop-corrected'' fall-offs.

In fact, one can argue that $A_{A}(u,\xh)$ obeys the same fall-offs as $u\ \rightarrow\ -\infty$. This is because, in a generic (classical) scattering,  charged particles accelerate under Coulombic drag in the infinte past as well. Due to this, early time radiation (in $(u,r,\hat{x})$ co-ordinates) has the same asymptotics as late time radiation. We thus see that Eq.(\ref{feb2-1})  can be extended to 
\be
A_A(u,\xh) \stackrel{u \to -\infty}{=} O(1) + O(1/u) + \cdots.
\ee
We also note that 
the coefficient $\overset{0,\ln}{F^+}_{rA}$ in (\ref{jan28-6}) is not an integration constant at $u\ =\ +\ \infty$ and hence 
can not be easily determined in terms of the asymptotic fields at time-infinity. In  appendix \ref{Frzapp} we show that using matter currents at time-infinity this mode can be written as 
\be\label{feb1-1}
 \overset{0,\ln}{F^+}_{rA}(\xh) = \frac{1}{8\pi}\ \int\ d^{3}y\ (-q\cdot Y)^{-3}L_{A}^{\mu\nu}(\hat{x})\ J_{\mu\nu}^{\alpha}(y)\ \overset{\ln}{j}_{\alpha}(y)
\ee
where $q^{\mu}\ =\ (1,\hat{x})$ and $J_{\mu \nu}^\alpha$ and $L^{\mu \nu}_A$ are angular momentum components defined at the asymptotic hyperboloid and celestial sphere respectively. See appendix \ref{Frzapp} for details.

\section{Revisiting the  subleading charges of QED with massive particles} \label{sec4}
In \cite{stromlow} it was shown that the subleading soft photon theorem was equivalent to a new class of infinite-dimensional symmetries parametrized by vector fields on the sphere. However this analysis was restricted to tree-level scattering amplitudes. Moreover the charged particles were  assumed to be massless. The generalization to massive particles was given recently in \cite{hirai}. In this section we review the construction of these ``subleading'' charges in the presence of massive particles.  Our discussion will be along the lines of \cite{subn}. 

The interest in the massive case is more then a mere technical complication as the loop-corrected $\log \w$ subleading soft photon theorem  trivializes for massless charged particles \cite{sahoo}.  Whence to understand the loop corrected soft theorem as Ward identities it is useful to first consider the tree-level case with massive charged particles. 





\subsection{Subleading charge} \label{ncharge}

Assuming the $O(e)$ fall-offs (\ref{Oefallu}), the (future) subleading charge defined in  \cite{stromlow} can be written as \cite{subn}\footnote{In \cite{subqed,subn} the charge is parametrized in terms of two scalars that smear the $O(u^0)$ coefficients of  $\overset{1}{F}_{ru}$ and its Hodge-dual $\overset{1}{\widetilde{F}}_{ru}$. The `purely electric' charge can be interpreted in terms of $O(r)$ large gauge transformations. The two descriptions are related through Maxwell equations: $\overset{1}{F}_{ru} = - D^{A}\overset{0}{F}_{rA}$, $\overset{1}{\widetilde{F}}_{ru} = \epsilon^{AB} D_A \overset{0}{F}_{rB}$.}
\be \label{stromcharge}
Q[V]= \int_{\scri^+_-} d^2 \xh V^A(\xh) \overset{0,1}{F^-}_{rA}(\xh)
\ee
where $\scri^+_-$ denotes the $u= -\infty$ sphere at future null infinity,   $V^A(\xh)$ is the vector field parametrizing the charge,  and $\overset{0,1}{F^-}_{rA}(\xh)$ is the $O(u^0)$ coefficient of $\overset{0}{F}_{rA}(u,\xh)$ at $u \to - \infty$, given in Eq. (\ref{OefalluFrA}).  

Analogous charge is defined at $\scri^-_+$ (the $v=+ \infty$ sphere at past null infinity) such that their equality is equivalent to the subleading soft photon theorem \cite{stromlow}. In \cite{subn} this equality was shown to be a consequence of field equations at spatial infinity under the assumption of  $O(e)$ fall-offs (\ref{Oefallu}) on both past and future null infinities.

The analysis of \cite{subn} was for the case of massless charged particles  and whence the matter contribution to the charge defined above was localised at null infinity. In this section we generalise  this derivation to our case of interest where the charged particles are massive. This will illustrate how tree-level (subleading) soft theorem is manifestation of Ward identities associated to the subleading charges defined in (\ref{stromcharge}).

Hence,  we want to write the charge $Q[V]$ as a sum of a soft charge and a hard charge, as in \cite{stromlow}. As the scattering particles are massive, we know that the hard charge will be localised at time-like infinity (i.e. at $u\ =\ \infty$) and the soft charge will be a flux at null infinity. More in detail, we aim to rewrite $Q[V]$ as 
\begin{equation}
Q[V]\ =\ \Qh[V]\ +\ \Qs[V]    
\end{equation}
where $\Qh[V]$ is integrated over the sphere at $u\ =\ \infty$ and $\Qs[V]$ is the soft charge integrated over ${\cal I}^{+}$. 
From Maxwell equations at null infinity  (\ref{leadingmaxwell}) one finds \cite{stromlow}
\be
\overset{0,1}{F^-}_{rA}(\xh)  = \lim_{u \to - \infty} \left[ \overset{0}{F}_{rA}(u,\xh) + u \big( \frac{1}{2} D_A \overset{0}{F}_{ru}(u,\xh) -   D^B D_{[A}A_{B]}(u,x)\ \big) \right] \label{strombdy}
\ee
The above expression can be trivially re-written as
\be
\overset{0,1}{F^-}_{rA}(\xh) =   \lim_{u \to + \infty} [\cdots] - \int_{-\infty}^\infty du \partial_u [\cdots]
\ee
where ``$[\cdots]$'' denotes the expression in square brackets in (\ref{strombdy}). The first term is just the $O(u^0)$ coefficient of $\overset{0}{F}_{rA}$ at $u \to \infty$:
\be
\lim_{u \to + \infty} [\cdots] = \overset{0,1}{F^+}_{rA}(\xh).
\ee
This yields the hard charge.\\
Using Maxwell equations (\ref{leadingmaxwell}), the integrand in the second term can be written as
\ba
-\partial_u [\cdots ] & = & u \partial_u\left( \frac{1}{2} D_A D^B A_B + D^B D_{[A}A_{B]} \right) .
\ea
The expression can be further simplified if written in terms of stereographic coordinates $(z,\zb)$. For $A=z$ it reduces to
\be
-\partial_u [\cdots] = u \partial_u D_z D^{\zb} A_{\zb}
\ee
Summarizing, the  charge (\ref{stromcharge}) can be written as
\be
Q[V] = \Qh[V] + \Qs[V]
\ee
where
\ba
\Qh[V] & = & \int_{\scri^+_+} d^2 \xh \, V^A \overset{0,1}{F^+}_{rA} \label{Qhardtree} \\
\Qs[V] & = & \int_{\scri^+} d^2 \xh du  \, V^z u \, \partial_u D_z D^{\zb}  A_{\zb} + (z \leftrightarrow \zb)
\ea

We finally would like to express (\ref{Qhardtree}) as an integral over $i^{+}$ involving the current. For this we use Eq. (\ref{bdytiA}) and  Maxwell's equation to express $\overset{1}{F}_{\tau A}$ in terms of the matter current as follows.

Let  $\G^\beta_\alpha(y;y')$ be the Green's function for the Poisson-type equation (\ref{poisson}),
\be
 \overset{1}{F}_{\t \alpha}(y) = \int d^3 y' \G_\alpha^\beta(y;y') \overset{0}{j}_\beta(y')\label{FnJn}
\ee
 and let $G_A^\beta(y'; \xh)$ be its leading  $\rho \to \infty$ asymptotic value,
 \be
 \G_A^\beta(\rho,\xh;y')  \overset{\rho \to \infty}{=} \rho^{-1} G_A^\beta(y'; \xh) + \ldots \label{Gnbdy} 
\ee
Then $\overset{1,0}{F}_{\tau A}$  can be written as
\be
\overset{1,0}{F}_{\tau A}(\xh)  = \int d^3 y'  \overset{0}{j}_\beta(y') G_A^\beta(y'; \xh).
\ee
We now use $G_A^\beta$ as a `boundary to bulk' Green's function for $V^A$ to define a hyperboloid vector field\footnote{Note that $G_{A}^{\beta}$ consists of ${\cal G}^{\rho}_{A}$ and ${\cal G}_{A}^{B}$ which collectively map the vector field on the celestial sphere to a vector field tangential to the asymptotic hyperboloid.}
 \be
V^\beta(y') := \int d^2 \xh \, G_A^\beta(y'; \xh) V^A(\xh)
\ee
In terms of this vector field, the charge (\ref{Qhardtree}) can finally be  written as
\be
\Qh[V] = \int d^3 y' V^\beta(y') \overset{0}{j}_\beta(y') .
\ee
In appendix \ref{greenapp} we evaluate $\G^\beta_\alpha(y;y')$ and $G_A^\beta(y'; \xh)$, and show that
\be
G_A^\beta(y; \qh) = \frac{1}{8 \pi}  (-q \cdot Y)^{- 3} J^\beta_{\mu \nu}(y) L^{\mu \nu}_A(\qh) 
\ee
where $ J^\beta_{\mu \nu}(y)$ and $L^{\mu \nu}_A(\qh)$ are the angular momentum generators on the hyperboloid and  sphere respectively. 

Thus, to summarise, the total charge $Q[V]$ at future infinity is 
\begin{equation}
Q[V]\vert_+ \ =\ \int_{i^+} d^3 y' V^\beta(y') \overset{0}{j}_\beta(y')\ +\ \int_{\scri^+} d^2 \xh du  \, V^z u \, \partial_u D_z D^{\zb}  A_{\zb} + (z \leftrightarrow \zb)
\end{equation}
We can write the charge at past infinity in exactly analogous fashion.

The  infinity of conservation laws parametrized by $V^{A}$ are then given by 

\begin{equation}
Q[V]\vert_{{\cal I}^{+}\cup\ i^{+}}\ =\ Q[V]\vert_{{\cal I}^{-}\cup\ i^{-}}
\end{equation}

In the next section we evaluate the corresponding Ward identities for tree-level S matrix and show that they are equivalent to subleading soft photon theorems (for both helicities) in scalar QED.

\subsection{Ward identities and tree level soft theorem}

We will for simplicity consider vector fields such that $V^{\zb}=0$ (corresponding to negative helicity subleading soft photon theorem).  In order to evaluate the S matrix Ward identity, we first need to express $Q[V]$ in terms of Fock operators.  For the soft charge, this is done by writing it in  frequency space as \cite{stromlow}
\be
\Qs[V] \ =\ \lim_{\omega\rightarrow 0}\partial_{\omega}\omega \int d^{2}z\ V^{z}(z,\zb)\ D_{z}^{2}A_{\zb}(\omega,z,\zb),
\ee
and then expressing $A_{\zb}(\omega,z,\zb)$ in terms of photon Fock operators.

As seen in the previous section, the hard charge is given by
\be
\Qh[V]\ =\ \ \int d^{3}y\ V^{\alpha}(y)\ \overset{0}{j}_{\alpha}(y)
\ee
with
\be
V^\alpha(y) = \frac{1}{8 \pi } \int d^2 \qh (-q \cdot Y)^{-3} J^\alpha_{\mu \nu}(y) L^{\mu \nu}_w(\qh) V^w(\qh). \label{Valpha}
\ee
Thus, to quantize $\Qh[V]$ we need to evaluate the asymptotic current $\overset{0}{j}_{\alpha}$ in terms of asymptotic fields.\footnote{We  remind the reader that  in this section we are considering $O(e)$ fall-offs of the fields 
and hence the matter field is taken to be free at late times.}
The $\t \to \infty$  asymptotics of the free massive field (\ref{freephi})
yields
\be
 \overset{0}{j}_\alpha = \frac{i e m}{4 (2 \pi)^3} (\ b \partial_\alpha b^* - b^* \partial_\alpha b\ ) - (\ b \leftrightarrow c\ ). \label{asymj0}
\ee
The quantum charge is then defined in terms of Fock operators $b, b^\dagger$, $c, c^\dagger$ by a normal ordered version of (\ref{asymj0}). From the commutator of the  Fock operators (written in hyperbolic coordinates)
\be
[b(y), b^\dagger(y')] = \frac{2 (2 \pi)^3}{m^2} \delta^{(3)}(y,y'), \label{fockcomm}
\ee
one immediately finds that the action of hard charge on scattering states can be evaluated using
\begin{equation}
\begin{array}{lll}
[b(y),\Qh[V]] = - i \frac{e}{m} V^\alpha(y) \partial_\alpha b(y) \\
\vspace*{0.1in}
[c(y),\Qh[V]] = + i \frac{e}{m} V^\alpha(y) \partial_\alpha c(y)
\end{array}
\end{equation}
 Thus, the Ward identity $Q[V]\vert_{+} S = S Q[V]\vert_{-}$ between given in and out states takes the form
\be
  \frac{1}{4 \pi }  \lim_{\w \to 0} \partial_\w \w \int d^2 w V^w D^2_w \frac{\sqrt{2}}{(1+ |w|^2)} \M_{N+1}(\w) =- \frac{e}{m} \sum V^\alpha \partial_\alpha \M_N \label{wardid}
\ee
where we have used  $A_{\wb}(\w,\qh) = \frac{1}{4 \pi i} \frac{\sqrt{2}}{(1+ |w|^2)}a_-(\w \qh)$ \cite{stromqed} and written the RHS for the case of positive outgoing charges (negative or incoming will have relative minus signs).

On the other hand, for tree level amplitudes, the subleading soft photon theorem can be written as
\be
\lim_{\w \to 0} \partial_\w ( \w \M_{N+1}(\w) ) = \sum S^{(1)}(q,y)  \M_N  \label{softthm}
\ee
where $S^{(1)}(q,y)$ is the differential operator 
\be
S^{(1)}(q,y)= \frac{e}{m} (q \cdot Y)^{-1} \e^\mu_- q^\nu J_{\mu \nu}^\alpha(y) \partial_\alpha  \label{S1}
\ee
(as in (\ref{wardid}), in (\ref{softthm}) we are also parametrizing the spatial components of hard momenta  by points $y^\alpha$ on the hyperboloid).

To go from the soft theorem (\ref{softthm}) to the Ward identity (\ref{wardid}) we smeared both sides of (\ref{softthm}) by
\be
 \frac{1}{4 \pi }  \int d^2 w V^w D^2_w \frac{\sqrt{2}}{(1+ |w|^2)} ,
\ee
so that the left hand sides coincide. The right hand side then coincide thanks to the identity\footnote{As a consistency check of the formulas, one can compute the $m \to 0$ limit of (\ref{keyid}) from Eq. (\ref{Gbdybdy}), by regarding the $m \to 0$ limit as $\rho = p/m \to \infty$ with $p=$const. Doing so one recovers the $m=0$ identitity $D^2_w\left( \frac{\sqrt{2}}{(1+ |w|^2)} S^{(1)}(q,p) \right) = \frac{2 \pi e}{p}   \delta^{(2)}(\xh,\qh) \partial_z+ \ldots$ which corresponds to Eq. (\ref{id2}). \label{fnmless}}
\be
D^2_w\left( \frac{\sqrt{2}}{(1+ |w|^2)} S^{(1)}(q,y) \right) = \frac{e}{2 m} \sqrt{\gamma} (q \cdot Y)^{-3} L^{\mu \nu}_w J^\alpha_{\mu \nu} \partial_\alpha \label{keyid}
\ee
To go in the reverse order, consider the kernel \cite{stromlow} 
\be
K_{\wb}^z :=  \frac{1}{2 \pi} \frac{(1+z \zb)}{(1+w \wb)}  \frac{(w-z)}{(\wb-\zb)}
\ee
which satisfies the identities
\ba
D^2_z K_{\wb}^z & = & \delta^{(2)}(w,z)  \label{id1} \\
D^2_w K_{\wb}^z  & = & \delta^{(2)}(w,z) \label{id2}
\ea
If we now take the Ward identity (\ref{wardid}) for the vector field 
\be
V^w : =  4 \pi  \frac{(1+ |w_0|^2)}{\sqrt{2}} K_{\wb_0}^w \label{VK}
\ee
then, by virtue of (\ref{id1}),  the LHS of (\ref{wardid}) becomes the LHS of (\ref{softthm}) for a soft photon with direction  $w_0$. Using (\ref{keyid}) to express the RHS of (\ref{wardid}) one easily recovers the RHS of (\ref{softthm}) thanks  again to (\ref{id1}).

\section{Defining $Q[V]$ with general fall-offs}

Now that we have warmed up by showing equivalence between tree-level subleading soft theorem and Ward identities associated to the subleading charges (which are conserved assuming $O(e)$ fall-offs), we extend the above definition of subleading charges and corresponding conservation laws for general fall-offs and quantize the corresponding charges. This will lead us to Ward identities which are equivalent to Sahoo-Sen subleading soft photon theorem.

In the previous section, the subleading charge 
was defined in terms of $\overset{0,1}{F^{-}}_{rA}(\hat{x})$ (Eq.  (\ref{stromcharge})). 
For the  general fall-offs we claim the  subleading charge 
is defined in terms of $\overset{0,\ln}{F^-}_{rA}(\xh) $, with  this coefficient  defined  in Eq.(\ref{jan28-6}).  We claim the charge  is conserved  in the sense that 
\be \label{lnconservation}
\overset{0,\ln}{F^-}_{rA}(\xh)\vert_{\scri^+} =\ \overset{0,\ln}{F^+}_{rA}(\xh)\vert_{\scri^-}
\ee
where $\overset{0,\ln}{F^+}_{rA}(\xh)\vert_{\scri^-}$ is the analogue coefficient at past null infinity. The conservation (\ref{lnconservation}) can be shown by studying the field equations at spatial infinity, along the same lines as in the case of the $O(e)$ fall-offs \cite{subn}. 

Our strategy for writing the charge as a sum of soft and hard charges is the same as  in the previous section. Whence we first rewrite $\overset{0,1}{F^{-}}_{rA}(\hat{x})$ as an integral over null infinity plus a boundary term at $u\ =\ +\infty$,
\be
\overset{0,\ln}{F^-}_{rA}(\xh) = - \lim_{u \to - \infty} u^2 \partial_u^2  \overset{0}{F}_{rA} = \int du \partial_u (u^2 \partial^2_u \overset{0}{F}_{rA}) + \overset{0,\ln}{F^+}_{rA}(\xh)  .
\ee
We will for simplicity focus on the $A=z$ component. The term at null infinity can be expressed in terms of the free data through the  field equations  
\be \label{FrzAz}
\partial_u^2  \overset{0}{F}_{rz} = -D_z D^{\zb} \partial_u A_{\zb}.
\ee
We then define 
\be
Q[V] :=  - \int d^2 \xh  V^z \overset{0,\ln}{F^-}_{rz}(\xh)  = :\Qs[V]+ \Qh[V].
\ee
The soft charge is given by\footnote{Here we used the identity $ \w \partial^2_\w \w =  \partial_\w \w^2 \partial_\w$.}
\be
\Qs[V] = \int d^2 z  V^z D^2_z \partial_u u^2 \partial_u A_{\zb} = \lim_{\w \to 0}  \partial_\w \w^2 \partial_\w  \int d^2 z  V^z D^2_z  A_{\zb}(\w,z,\zb)
\ee
This has the same angular form as the tree-level soft charge, but with the $\w$ factors now isolating a $O(\ln \w)$  coefficient in the $\omega$ expansion!\footnote{More in detail: $\lim_{\omega\rightarrow 0}\partial_{\omega}\omega^{2}\partial_{\omega}\ (\frac{\overset{-1}{f}}{\omega}\ +\ \ln\omega\ \overset{\ln}{f}\ + O(\omega^{0}) )\ =\ \overset{\ln}{f}$}
The hard charge is given by
\be
\Qh[V] = - \int d^2 \xh  V^z \overset{0,\ln}{F^+}_{rz}(\xh), \label{Qhard}
\ee
which, using Eqs. (\ref{feb1-1}) and (\ref{Valpha}) can be written as 
\be
\Qh[V] = - \int d^3 y V^\alpha(y) \overset{\ln}{j}_\alpha(y)
\ee
with $V^\alpha(y)$ defined as before, Eq. (\ref{Valpha}).

We finally discuss the form of $\overset{\ln}{j}_\alpha(y)$ in terms of the asymptotic Fock operators.  From Eq.(\ref{dressedphi}) we have 
\be
j_\alpha(\t,y)  \stackrel{\t \to \infty}{=} \ln \t 2 e^2 \partial_\alpha A_\t(y) |\phifree(\t,y)|^2
\ee
Using the expression (\ref{freephi}) for the free field in terms of Fock operators we then find
\ba
\overset{\ln}{j}_\alpha(y) & = & \frac{m e^2}{2(2\pi)^3} [b^\dagger(y) b(y) + c^\dagger(y) c(y) ] \partial_\alpha A_\t(y)\\
& = & \frac{e^2}{m} n(y)\partial_{\alpha}A_\t(y)  
\ea
where $n(y)\ =\ \frac{m^2}{2(2\pi)^{3}}\ [b^\dagger(y) b(y) + c^\dagger(y) c(y) ]$ is the number density operator  and $ A_\t(y)$ is  a sum of two terms
\be
 A_\t(y) =  A^j_\t(y) +  A^\hom_\t(y)
\ee
such that
\be
\D^2  A^j_\t = - \overset{0}{j}_\t, \quad \quad  \D^2 A^\hom_\t=0.
\ee 
Thus $A^j_\t(y)$ is the asymptotic scalar potential sourced by the charged matter current and 
$A^\hom_\t$ is a homogeneous piece that, as we will see depends on the soft mode of the photon field.

We first analyse $A^j_\t(y)$. 
By solving the  asymptotic Maxwell equations one can write  $\overset{0}{A}_\t$ in terms of $\overset{0}{j}_\t$ as \cite{grey}:
\be\label{grey}
A^j_\t(y) = \frac{1}{4 \pi} \int d^3 y' \big(\frac{Y \cdot Y'}{\sqrt{(Y \cdot Y')^2-1}} +1 \big) \overset{0}{j}_\t(y'), 
\ee
where $\overset{0}{j}_\t$ is  given by
\be
\overset{0}{j}_\t(y')= \frac{m^2 e}{2 (2 \pi)^3} [ - b^\dagger(y') b(y') + c^\dagger(y') c(y') ] .
\ee
We now analyse $A^\hom_\t(y)$.\footnote{An alternative description of  $A^\hom_\t(y)$ in terms of Kulish-Faddeev fields is given in section \ref{KFsec}.} Understanding this term is subtle since  we expect  the homogeneous part of the scalar potential ${\cal A}_{\tau}$ 
falls  as $\frac{1}{\tau^{2}}$ implying $A^\hom_\t(y)=0$. 
As we show below, a detailed analysis of $A^\hom_\t(y)$ reveals that it is fixed by a  ``mode" of $A_{A}(u,\hat{x})$ which behaves as $\ln u\ \overset{\ln}{A}_{A}(\hat{x})$. However, as can be seen from Eq. (\ref{feb2-1}), this mode is absent in the classical theory (the presence of such term would yield an infinite angular momentum flux \cite{herdegen}). 
Rather remarkably  this term turns out to be non-trivial 
in quantum theory and contributes to the quantized charge.

In order to discuss the relation between the $u \to \infty$ and $\t \to \infty$  asymptotics it is convenient to work in a representation of $\A_\mu$ used by Herdegen \cite{herdegen}: A general solution of free Maxwell equations in Lorentz gauge may be written as\\
 
\ba
{\cal A}_\mu(x) = - \frac{1}{2 \pi} \int d^2 \qh \, \Omega^{-1} D^A(\Omega q_\mu) \dot{A}_A(- q \cdot x,\qh), \label{herdegen}
\ea
where $\dot{A}_A(u,\qh) \equiv \partial_u A_A(u,\qh)$ is the $u$-derivative of the radiative data at future null infinity,  $q^\mu=(1,\qh)$, and $\Omega = \Omega(\qh) = (1 + z\zb)/\sqrt{2}$.\footnote{Herdegen gives a more generic expression for ${\cal A}_{\mu}(x)$ parametrized by different choices of (residual) gauge transformations. We have fixed the residual gauge and chosen a given $\Omega$ such that Fourier transform of ${\cal A}_{\mu}(x)$ matches with the standard expressions in Lorenz gauge \cite{strominger}. 
See  section \ref{KFsec} for further details.}

Expression   (\ref{herdegen}) can now be used to relate the $u \to \infty$ with $\t \to \infty$ asymptotics in a straightforward manner. 

It says that the $\t \to \infty$ decay of ${\cal A}_\mu$ is dictated by the $u \to \infty$ decay of $\partial_u A_A(u,\qh)$. Here we note that if our radiative data were as in Eq.(\ref{feb2-1}), we would have $\partial_u A_A(u,\qh) = O(1/u^2)$ and so ${\cal A}_\mu=O(1/\t^2)$ as expected. Whence a $O(\t^{-1})$  term is associated to a  $\log u$ piece in the radiative data: 
\be
A_A(u,\xh) \stackrel{u \to \infty}{=} \overset{\ln}{A}_A(\xh) \log u + \ldots \label{Alnu}
\ee
The time-infinity contribution from such logarithmic term can be obtained by substituting (\ref{Alnu}) in (\ref{herdegen}) and taking $x = \t Y$. For ${\cal A}_\t \equiv Y^\mu {\cal A}_\mu$ we get

\be
{\cal A}_\t =  -\frac{1}{2 \pi} \int d^2 \qh \, \Omega^{-1}  D^A(\Omega Y \cdot q) \frac{\overset{\ln}{A}_A(\qh)}{(-\t Y \cdot q)} + \ldots =  \frac{1}{\t} A^\hom_\t  + \ldots
\ee
with
\be\label{feb2-4}
A^\hom_\t = - \frac{1}{2 \pi} \int d^2 \qh \,  \log( - Y \cdot q) D^A \overset{\ln}{A}_A(\qh) \ + \ \text{($y$-independent)}
\ee
where the $y$-independent term is $- \frac{1}{2 \pi} \int d^2 \qh \ln (\Omega) D^A \overset{\ln}{A}_A$.  In the following we will ignore such  term since it is irrelevant for  the current $ \overset{\ln}{j}_\alpha \sim  \partial_\alpha A_\t$.


Whence the total hard charge $\Qh[V]$ is given by

\begin{equation}
\Qh[V]\ :=\ \Qh_{j}[V]\ +\ \Qh_{\hom}[V]
\end{equation}
with
\begin{equation}\label{Qhard2}
\begin{array}{lll}
\Qh_{j}[V]\ =\ -\int d^{3}y n(y)\  V^{\alpha}(y)\partial_{\alpha} A_{\tau}^{j}(y)\\
\vspace*{0.1in}
\Qh_{\hom}[V]\ =\ -\int d^{3}y n(y) V^{\alpha}(y)\ \partial_{\alpha} A_{\tau}^{\hom}(y)
\end{array}
\end{equation}

In the next section we quantize the hard charge $\Qh[V]$. We conclude this section by noting that 
\begin{itemize}
\item Quantization of $\Qh_j[V]$ in Eq.(\ref{Qhard2}) is rather straightforward as it is a quartic operator in charged scalar field as can be seen from Eq. (\ref{grey}).

\item As we will see in the next section and in contrast to the tree level hard charge, the quantum $\Qh[V]$ has a non-trivial contribution  from the gauge field mode at $u\ =\ +\infty$, encoded in the non-trivial quantum operator corresponding to $\Qh_\hom[V]$ .
\end{itemize}

\subsection{{ Deriving  $\Qh_\hom[V]$}}

We now discuss the quantization of  $\Qh_{\hom}[V]$. As is clear from Eq.(\ref{Qhard2}), this entails quantizing  $\overset{\ln}{A}_A(\xh)$. Before going into the technicalities, let us summarise the final result:   $\overset{\ln}{A}_A(\xh)$ is given by the  linear combination of soft Fourier modes,
\begin{equation}\label{feb2-5}
\overset{\ln}{A}_z=  \frac{i}{8 \pi^2} \frac{\sqrt{2}}{1+z \zb} \lim_{\w \to 0^+} \w( a_+(\w, \xh) - a^\dagger_-(\w,\xh)) .
\end{equation}

We start with some observations regarding the relation between $|u| \to \infty$ and $|\w| \to 0$ fall-offs for a class of functions. Consider a function $f(u)$ with Fourier transform
\be
\tilde{f}(\w) = \int f(u) e^{i \w u } du.
\ee
If $f(u)$ has the $u \to \pm \infty$ asymptotics
\be
f(u) \stackrel{u \to \pm \infty}{=} c^\pm + \cdots
\ee
then its Fourier transform satisfies\footnote{One way to see the validity of (\ref{tildef}) is to write
$f(u) \stackrel{|u| \to \infty}{=} c^+ \theta(u) + c^- \theta(-u) + \cdots$
where $\theta(u)$ is the step function.  The first two terms in (\ref{tildef}) are  the Fourier transforms of $\theta(\pm u)$.}
\ba
\tilde{f}(\w) & \stackrel{\w \to 0}{=} &  c^+  \frac{i}{\w + i \epsilon}  - c^- \frac{i}{\w - i \epsilon}  + \cdots \label{tildef} \\
&=& (c^+ - c^-) i P\frac{1}{\w} + (c^+ + c^-) \pi \delta(\w) +\cdots \label{tildef2}
\ea
In all expressions the dots denote subleading terms whose precise form will not matter for the present discussion. (For the radiative data of interest the subleading term in $u$ is $O(1/u)$ corresponding to a  $O(\log \w)$ term in frequency \cite{observational}.) 

From (\ref{tildef2}) we see we can write
\be
c^+-c^- = -i \lim_{\w \to 0} \w \tilde{f}(\w) , \label{limwf}
\ee
Note that in Eq. (\ref{limwf}) it does not matter whether   $\w$ approaches zero from above or below. 
In quantum theory however, the $\w \to \pm 0$ limits are different since they involve  either creation of annihilation operators.

We now consider the positive/negative frequency decomposition for $f(u)$:
\be
f(u)= f^{+}(u) + f^-(u)
\ee
where
\be
f^+(u) = \int_0^\infty \frac{d \w}{ 2 \pi} \tilde{f}(\w) e^{ - i \w u}, \quad  f^-(u)= (f^+(u))^*
\ee
and ask what is the $u \to +\infty$ asymptotics of $f^\pm(u)$. This can be evaluated by considering the Fourier transform of $\partial_u f^\pm$ and then integrating. One finds
\be
f^\pm(u) \stackrel{u \to +  \infty}{=}  \frac{c^+}{2} \pm \frac{\log u}{2 \pi i} (c^+-c^-) + \ldots 
\ee
where the dots can include a purely imaginary $u^0$ piece plus subleading terms. 
We can now express the $\log u $ coefficient by means of Eq. (\ref{limwf}). Since  $f^\pm(u)$  only knows about  positive/negative values of $\w$, this coefficient is related to the ``soft Fourier mode" via, 
\be\label{feb12-3}
c^+-c^- = -i \lim_{w \to 0^\pm} \w \tilde{f}(\w).
\ee
This can be seen as follows. Consider  
\begin{equation}
\begin{array}{lll}
\partial_{u}f^{+}(u)\ &&\stackrel{u\ \to\ +\infty}{=} \frac{1}{2\pi i u}(c^{+} - c^{-})\ +\ \dots\\
\end{array}
\end{equation}

Using Fourier transform we also know that 

\begin{equation}
\begin{array}{lll}
\partial_{u}f^{+}(u)\ &&=\ \frac{1}{2\pi i}\ \int d\omega\ (\omega\tilde{f}(\omega)) \ e^{-i\omega u}\\
\vspace*{0.1in}
&&=\ \frac{1}{+ 2\pi  u}\int d\omega \ (\omega\tilde{f}(\omega)) \frac{d}{d\omega}e^{-i\omega u}\\
\vspace*{0.1in}
&&=\ \frac{1}{+2\pi  u}\left[\ \int d\omega \frac{d}{d\omega}(\ \omega\tilde{f}(\omega) e^{-i\omega u}\ )\ -\ \int d\omega \frac{d}{d\omega}(\omega\tilde{f}(\omega)\ e^{-i\omega u} \right]
\end{array}
\end{equation}

Thus in $u\ \rightarrow\ \infty$ limit, 

\begin{equation}\label{feb12-1}
\begin{array}{lll}
\frac{1}{2\pi i u}(c^{+} - c^{-})\ &&\stackrel{u \to + \infty}{=}\ \frac{1}{+ 2\pi  u}\left[\ \int d\omega \frac{d}{d\omega}(\omega\tilde{f}(\omega) e^{-i\omega u})\ -\ \int d\omega \frac{d}{d\omega}(\omega\tilde{f}(\omega))\ e^{-i\omega u}\right]\\
\vspace*{0.1in}
&& = -\frac{1}{2\pi u}\lim_{\omega\rightarrow 0^{+}}\omega\tilde{f}(\omega)
\end{array}
\end{equation}
Thus we see that starting from $f^{+}(u)$, we have 

\begin{equation}
(\ c^{+}\ -\ c^{-}\ ) \ =\ -i\ \lim_{\omega\rightarrow 0}\omega\ \tilde{f}(\omega)  
\end{equation}

We can similarly show that starting from $f^{-}(u)$, one gets,

\begin{equation}\label{feb12-2}
c^{+}\ -\ c^{-}\ =\ -i\ \lim_{\omega\rightarrow 0^{-}}\omega\tilde{f}(\omega)
\end{equation}

This proves Eq.(\ref{feb12-3}). 

Writing $f(u)= f^{+}(u) + f^-(u)$ we finally find
\be
f(u)  \stackrel{u \to + \infty}{=} c^+ +  \log u \overset{\ln}{f} + \ldots
\ee
with
\be
\overset{\ln}{f} = - \frac{1}{2 \pi} \lim_{\w \to 0^+} \w(\tilde{f}(\w) +\tilde{f}(-\w) ) . \label{fln}
\ee
By applying the above considerations to $A_A(u,\xh)$ and using the relation between $\tilde{A}_A(\w,\xh)$ and the photon Fock operators \cite{strominger}
\be \label{fourier1d}
\tilde{A}_z(\w,\xh) = \left\{
\begin{array}{cl}
\frac{1}{4 \pi i} \frac{\sqrt{2}}{1+|z|^2} a_+(\w, \xh)  & \text{for } \quad \w>0 \\
\\
-\frac{1}{4 \pi i} \frac{\sqrt{2}}{1+|z|^2} a^\dagger_-(-\w, \xh)  & \text{for } \quad \w < 0 
\end{array}\right.
\ee
one arrives at  Eq.(\ref{feb2-5}). $\overset{\ln}{A}_{\zb}$ is given by a similar expression with $+$ and $-$ interchanged. 


For later purposes, it will be useful to rewrite the  expression for $\overset{\ln}{A}_A$ as 
\be
\overset{\ln}{A}_{A}  = \frac{1}{2 \pi i}([A_{A}]^+ - [A_{A}]^-) \label{fln2}
\ee
where
\be \label{defsqbA}
[A_{A}(\hat{x})]^\pm =  -i \lim_{\w \to 0^\pm} \w \tilde{A}_{A}(\w,\hat{x})
\ee
is the evaluation of $[A_{A}]$ using only positive/negative frequencies.\footnote{That is, $\lim_{\w \to 0^{+}} \w \tilde{A}_{z}(\w)\ \sim\ \lim_{\w \to 0}\w a_{-}(\w)$ and $\lim_{\w \to 0^{-}} \w \tilde{A}_{z}(\w)\ \sim\ -\lim_{\w \to 0}\w a_{+}^{\dagger}(\w)$}
In terms of these quantities, we can express Eq. (\ref{feb2-4}) as
\be
A^\hom_\t(y)  = - \frac{1}{4 \pi^2 i} \int d^2 \qh \,  \log( - Y \cdot q)( D^A [A_A(\qh)]^+ - D^A [A_A(\qh)]^-)  .\label{Ahom2} 
\ee
 
\subsection{Quantization of the hard charge}

Having derived the quantum operator associated to $\overset{\ln}{A}_{z}$ (or more generally $\overset{\ln}{A}_{A}$), we are in a position to quantize $\Qh[V]$, which as we recall, is given by
\begin{equation}
\Qh[V]\ =\ - \frac{e^2}{m} \int d^{3}y\ n(y)\ V^{\alpha}(y)\ \partial_{\alpha} A_\t(y)
\end{equation}
where
\be
n(y)  =  \frac{m^2}{2(2\pi)^{3}}\ [b^\dagger(y) b(y) + c^\dagger(y) c(y) ] \nn 
\ee
is the number-density operator 
and
\be
A_\t(y) = A_{\tau}^{j}(y)\ +\ A_{\tau}^{\hom}(y)
\ee
given in Eqs.(\ref{grey}), (\ref{feb2-4}).
Let 
\be
| \out \ket = | (e_1,y_1) , (e_2, y_2), \cdots \ket
\ee
be an `out' state consisting of particles with momenta labeled by  $y_1,y_2,\cdots$ and electric charge $e_1, e_2, \cdots$ where $e_i = \pm e$ for a  particle/antiparticle.  The action of $n(y)$ on such state is easly obtained from the Fock commutation relation:
\be
n(y) | \out \ket = \sum_{a \in \out} \delta^{(3)}(y,y_a) | \out \ket,
\ee
from which we find\footnote{There is a subtelty here in that $n(y)$ and $A^j_\t(y)$ do note commute. Thus, strictly speaking, Eq. (\ref{actionQh}) is only valid for an operator ordering where $n(y)$ appears to the right of  $A^j_\t(y)$. The action for the corrected, normal ordered operator, can however be easily obtained  from Eq. (\ref{actionQh}), see Eq. (\ref{QjoutS}) and footnote \ref{fnotenormal}.}
\be \label{actionQh}
\Qh | \out \ket  = - \frac{e^2}{m}  \sum_{a \in \out} V^\alpha(y_a) \frac{\partial}{\partial y^\alpha_i} A_\t(y_a) | \out \ket .
\ee
We next need to compute the action of  $A_\t(y_i)$ on the out state. Rather than doing so directly, we will consider the computation within  S matrix elements,
\be \label{actionQhS}
\bra \out | \Qh S | \inn \ket =  - \frac{e^2}{m}  \sum_{a \in \out} V^\alpha(y_a) \frac{\partial}{\partial y^\alpha_a} \bra \out | A_\t(y_a) S | \inn \ket ,
\ee
as this  will bring out simplifications and will suffice for the evaluation of Ward identities.  
We write
\be
\bra \out | \Qh S | \inn \ket  = \bra \out | \Qh_{j}S | \inn \ket +  \bra \out | \Qh_{\hom} S | \inn \ket
\ee
to denote the contributions from  $ A_{\tau}^{j}$ and  $A_{\tau}^{\hom}$ respectively. The first contribution can  be evaluated from  the action of $\overset{0}{j}_\t(y')$ on the out state,
\be
\overset{0}{j}_\t(y') | \out \ket = - \sum_{a \in \out} e_a \delta^{(3)}(y',y_a) | \out \ket
\ee
together with  Eq. (\ref{grey})
\be
A^{j}_\t(y)  = \frac{1}{4 \pi} \int d^3 y' \big(\frac{Y \cdot Y'}{\sqrt{(Y \cdot Y')^2-1}} +1 \big) \overset{0}{j}_\t(y'), \label{grey2}
\ee
leading to\footnote{The correct, normal ordered charge yields Eq. (\ref{QjoutS}) with the divergent  $j=i$ terms absent.  We will implicitly assume the restriction $j \neq i$  on the double sum.  \label{fnotenormal}}
\be \label{QjoutS}
\bra \out | \Qh_{j} S | \inn \ket =  \frac{e^{2}}{4 \pi m}   \sum_{a,b \in \out} e_{b}\ V^\alpha(y_a) \frac{\partial}{\partial y^\alpha_a} \left( \frac{ Y_a \cdot Y_b}{\sqrt{(Y_a \cdot Y_b)^2-1}} \right) \bra  \out |  S | \inn  \ket.
\ee


We now evaluate the second contribution,
\ba
\int d^{3}y\ V^{\alpha}(y)\ \bra  \out | n(y)   \partial_\alpha A^\hom_\t(y) S | \inn \ket \nonumber\\
& = &  - \frac{e^2}{m} \sum_{a \in \out} V^\alpha(y_a) \frac{\partial}{\partial y^\alpha_a} \bra \out| A^\hom_\t(y_a) S | \inn  \ket\nonumber\\ \label{Q2outeval2}
\ea
The action of $A^\hom_\t$  in the above equation can be evaluated by noting that (\ref{Ahom2}) is closely related to the leading soft charge \cite{stromqed}. 
\be
\Qs_0[\lambda] := \frac{1}{2} \int d^2 \qh \lambda(\qh) ( D^A [A_A(\qh)]^+ + D^A [A_A(\qh)]^-).
\ee
Let us see why this is the case. Consider a specific choice of the celestial function $\lambda$ given by, 
\be 
\lambda_a(\qh) := \log (-q \cdot Y_a).
\ee
Taking this function as a  (leading) large gauge parameter,\footnote{Note that the sub-script $a$ indicates a point on the hyperboloid and is not a co-vector index}  the corresponding  gauge parameter at time infinity is given by \cite{massive}
\be \label{Lam_a}
\Lambda_a(y) := \frac{1}{4 \pi} \int d^2 \qh \frac{\lambda_a(\qh)}{(q \cdot Y)^2}.
\ee
 We then have
\ba\label{feb9-7}
 \bra \out| A^\hom_\t(y_a) S | \inn  \ket & = &  - \frac{1}{4 \pi^2 i}\bra \out| 2 \Qs_0[\lambda_a] S | \inn \ket \nonumber\\
& = &  - \frac{1}{4 \pi^2 i} \sum_{b} e_b \Lambda_a(y_b)
\ea
where we used the leading Ward identity together with $\Qs_0[\lambda] S = - S \Qs_0[\lambda]$. The integral (\ref{Lam_a}) can be shown to be given by 
\be
\Lambda_a(y)=  \frac{1}{2} \frac{Y \cdot Y_a}{\sqrt{Y \cdot Y_a^2-1}} \log \left( \frac{Y \cdot Y_a+ \sqrt{Y \cdot Y_a^2-1}}{Y \cdot Y_a- \sqrt{Y \cdot Y_a^2-1}} \right)+ \text{ terms independent of }  y_a . \label{Lambda}
\ee
Substituting eqn. (\ref{feb9-7}) in RHS of eqn. (\ref{Q2outeval2}) we get,
\begin{equation}\label{feb6-2}
\langle\textrm{out}\vert \Qh_{\hom}[V]\ S\ \vert\textrm{in}\rangle\ =\ \frac{e^{2}}{4\pi^{2} i m}\sum_{a\in\textrm{out}, b}\ e_{b}\ V^{\alpha}(y_{a})\partial_{\alpha}\Lambda_{a}(y_{b})\ \langle\textrm{out}\vert\ S\ \vert\textrm{in}\rangle
\end{equation}

We can similarly evaluate $\langle\textrm{out}\vert\ S\ \Qh[V] \vert\textrm{in}\rangle$ and combining the result with  eqns. (\ref{QjoutS}, \ref{feb6-2} ) we get,

\begin{equation}
\begin{array}{lll}
\langle\textrm{out}\vert\ [\Qh[V],\ S]\ \vert\textrm{in}\rangle\ =\\
\vspace*{0.1in}

\frac{e^{2}}{4 \pi m} \left( \sum_{a,b \in \out} e_{b}\  V^\alpha(y_a) \frac{\partial}{\partial y^\alpha_a} \left( \frac{ Y_a \cdot Y_b}{\sqrt{(Y_a \cdot Y_b)^2-1}} \right)\ -\ \out \leftrightarrow \inn\right)\\
\vspace*{0.1in}
\frac{e^{2}}{4\pi i m}\left(\sum_{a\in \out, b}\  -\ \sum_{a\in\ \inn, b}\right) e_{b}\ V^{\alpha}(y_{a})\partial_{\alpha}\Lambda_{a}(y_{b})\   \\
\vspace*{0.1in}
\hspace*{3.5in} \bra  \out |  S | \inn  \ket.
\end{array}
\end{equation}

We can re-express the RHS of the above equation in a more compact form as follows. Let 

\begin{equation}\label{s1}
\begin{array}{lll}
s_{1}(Y_{a}, Y_{b})\ &&=\ +\ e^{2}e_{b}\frac{ Y_a \cdot Y_b}{\sqrt{(Y_a \cdot Y_b)^2-1}}\ \textrm{if}\ (a,b)\ \in\ \out\\
\vspace*{0.1in}
&&= -\ e^{2}e_{b}\frac{ Y_a \cdot Y_b}{\sqrt{(Y_a \cdot Y_b)^2-1}}\ \textrm{if}\ (a,b)\ \in\ \inn\\
\vspace*{0.1in}
&&=\ 0\ \textrm{otherwise}
\end{array}
\end{equation}

and 

\begin{equation}\label{s2}
\begin{array}{lll}
s_{2}(Y_{a}, Y_{b})\ &&=\ - i\ e_{b}\ \Lambda_{a}(Y_{b})\ \forall\ a\ \in \out\\
\vspace*{0.1in}
&&=  i\ e_{b}\  \Lambda_{a}(Y_{b})\ \forall\ a \in \inn
\end{array}
\end{equation}

We then have

\begin{equation}\label{feb11-1}
\begin{array}{lll}
\langle\textrm{out}\vert\ [\Qh[V],\ S]\ \vert\textrm{in}\rangle\ =\\
\vspace*{0.1in}
\frac{e^{2}}{4 \pi m}\sum_{a,b}V^{\alpha}(y_{a})\frac{\partial}{\partial y_{a}^{\alpha}}\left(s_{1}(Y_{a}, Y_{b})\ +\ s_{2}(Y_{a}, Y_{b})\right)\  \langle\textrm{out}\vert S\vert\textrm{in}\rangle
\end{array}
\end{equation}

\section{Ward identity and Sahoo-Sen subleading theorem}

Let $S\ =\ \sum_{n=0}^{\infty} S_{(n)}$ where $\sum_{n=0}^{N}S_{(n)}$ denotes  the (loop-corrected) S matrix operator up to order $O(e^{2N})$ .  $S_{(0)}$ is the tree-level S matrix. 
The ward identity for $Q[V]$ can then be written in a loop expansion as, 
\begin{equation}
\begin{array}{lll}
\langle \out\vert [\ \Qs[V],\ S_{(n)}\ ]\vert \inn\rangle\ =\ -\langle \out\vert [\ \Qh[V],\ S_{(n-1)}\ ]\vert \inn\rangle\
\end{array}
\end{equation}

Whence the Quantum-corrected Ward identity (for the sub-leading charge) in scalar QED is given by,

\begin{equation}
\begin{array}{lll}
\langle \out\vert [\ \Qs[V],\ S_{(n)}\ ]\vert \inn\rangle\ =\ -\ \frac{1}{4 \pi m}\sum_{a,b}V^{\alpha}(y_{a})\frac{\partial}{\partial y_{a}^{\alpha}}\left(s_{1}(Y_{a}, Y_{b})\ +\ s_{2}(Y_{a}, Y_{b})\right)\  \langle\textrm{out}\vert S_{(n-1)}\vert\textrm{in}\rangle
\end{array}    
\end{equation}

We now claim that this Ward identity is equivalent to the Sahoo-Sen sub-leading Soft photon theorem in Scalar QED. 

Let us for simplicity, specialise to 1-loop S matrix. In this case, the Ward identity gives ,

\begin{equation}
\begin{array}{lll}
\langle \out\vert [\ \Qs[V],\ S_{t} ]\vert \inn\rangle\ =\ 0\\
\vspace*{0.1in}
\langle \out\vert [\ \Qs[V],\ S_{l}\ ]\vert \inn\rangle\ =\ -\langle \out\vert [\ \Qh[V],\ S_{t}\ ]\vert \inn\rangle\
\end{array}
\end{equation}

The first  of the above two equations can be re-written as, 
\begin{equation}
\lim_{\omega\rightarrow 0}\partial_{\omega}\omega^{2}\partial_{\omega}{\cal M}_{n+1}^{\textrm{tree}}(p_{1},\dots, p_{n},k)\ =\ 0
\end{equation}
which is consistent with the tree-level soft photon theorem which has no $\ln\omega$ term.

The second equation yields a ``1-loop corrected" Ward identity, which using eqn.(\ref{feb11-1}) can be written as 

\begin{equation}\label{feb12-6}
\langle \out\vert [\ \Qs[V],\ S_{l}\ ]\vert \inn\rangle\ =\ -\ \frac{e^{2}}{4 \pi m}\sum_{a,b}V^{\alpha}(y_{a})\frac{\partial}{\partial y_{a}^{\alpha}}\left(s_{1}(Y_{a}, Y_{b})\ +\ s_{2}(Y_{a}, Y_{b})\right)\  \langle\textrm{out}\vert S_{t}\vert\textrm{in}\rangle
\end{equation}

We will now show how this Ward identity is equivalent to the logarithmic sub-leading soft photon theorem for 1-loop QED scattering amplitude. 

\begin{equation}\label{ss-soft}
\lim_{\omega\rightarrow 0}\partial_{\omega}\omega^{2}\partial_{\omega}\ {\cal M}^{l}_{n+1}(\textrm{out},\textrm{in},k)\ =\ \sum_{a,b}S^{(\ln)}(p_{a},p_{b})\ {\cal M}_{n}^{t}(\textrm{out},\ \textrm{in})
\end{equation}

The LHS of eqn.(\ref{ss-soft}) contains a 1-loop corrected scattering amplitude which is infra-red divergent and is regularised in \cite{sahoo} using the technique developed by Grammer-Yennie in \cite{gy}. The RHS contains a logarithmic soft factor multiplying tree-level scattering amplitude given in eqn. (\ref{ss-log}). Thus formally, the soft theorem has the same structure as the sub-leading Ward identity. We now show that they are equivalent. 
Following the same steps leading to the  tree-level Ward identity (\ref{wardid}), we arrive at the following form for eqn. (\ref{feb12-6}).
\be
  \frac{1}{4 \pi}  \lim_{\w \to 0} \partial_\w \w^2 \partial_\w \int d^2 w V^w D^2_w \frac{\sqrt{2}}{(1+ |w|^2)} \M_{N+1}(\w) =-  i  \frac{e^2}{m} \sum_{a,b \in \out}  e_b s(y_a, y_b) \M_N - (\out \leftrightarrow \inn) \label{wardlnw}
\ee
To obtain the soft theorem, we choose $V^w$ as in Eq. (\ref{VK}) so that the LHS of (\ref{wardlnw}) becomes
\be
LHS=  \lim_{\w \to 0}  \partial_\w \w^2 \partial_\w \M_{N+1}(p_1,\ldots ;(\w \qh(w_0,\wb_0), \e_-)).
\ee
On the other hand, for this choice of $V^w$ we have
\be
V^\alpha \partial_\alpha = - (q \cdot Y)^{-1} \e^\mu_- q^\nu J_{\mu \nu}^\alpha(y) \partial_\alpha  ,
\ee
which was the identity that allowed us to recover the subleading tree-level soft theorem from the Ward identity. Going back to the definition for $s_{1}(y,y')\ +\ s_{2}(y,y')$ given in eqns.(\ref{s1}) and  (\ref{s2}) we can readily show that  the RHS of (\ref{wardlnw}) is precisely given by RHS of eqn. (\ref{ss-soft}). 

Several comments are in order.

\begin{itemize}
\item Rather remarkably, the quantum piece in the log corrected soft photon theorem which includes sum over all pairs of particles naturally arises due to a Quantum correction to the hard charge. In \cite{sahoo}, it was shown that this quantum piece is small when  back-reaction effect of emitted radiation on scattering particles can be neglected. We see that even from the perspective of Asymptotic charge, the quantum correction to the hard charge is small when particles suffer small deflection. This is due to the fact that this hard charge is governed by leading soft photon factor which vanishes in the limit of vanishing deflection of scattering particles.

\item We recall that the central beast in our derivation of $\Qh[V]$ is $A_{\tau}\ =\ A_{\tau}^{j}\ +\ A_{\tau}^{\hom}$. As we show in section \ref{KFsec}, $A_{\tau}$ has a clear relationship with Kulish-Faddeev Dressing factor. 
\item As the hard charge is quartic in matter operators, The algebra generated by $Q[V]$ is not understood and hence we so far lack any understanding of Asymptotic symmetry associated to the conservation laws. 
\end{itemize}

\section{Relationship with KF dressing}\label{KFsec}
One of the key concerns regarding symmetries of S-matrix in QED and Quantum Gravity comes from the fact that in four dimensions due to infra-red divergences the S-matrix does not exist and is only a formally defined object. Of course, as soft theorem is really a statement concerning ratios of two Scattering amplitudes (involving amplitudes with and without soft particle), it is a well defined statement even in the absence of a well-defined S matrix \cite{sahoo}. However when it comes to formulating statements about symmetries of S-matrix, we need a well defined S matrix!\\
If we restrict our analysis to tree-level scattering amplitudes, this issue does not arise as there are no IR divergences  but in the current example, when we are trying to understand loop corrected soft theorems as Ward identities, this issue becomes relevant.   As is well known, in the case of QED (and Gravity) one can in fact define infra-red finite S matrix with more careful treatment of the charged scattering states which acquire a dressing due to long range Coulombic effect. 
Whence the question we would like to ask is, if the Ward identities associated to the conservation laws are equivalent to Sahoo-Sen subleading theorem when evaluated in dressed states. Although we do not answer this question in this paper, we show how the above mentioned dressing is directly related to the Asymptotic matter fields at time-infinity. This observation may be of some relevance when analysing these Asymptotic symmetries in dressed states.

We now show that the dressing of the scalar field described in the paper can be recovered from the analysis  of  Kulish and Faddeev  \cite{KF}.  We will partly follow the KF treatment as presented in \cite{akhouryKF}.

In the KF formulation,  the creation and annihilation Fock operators associated to the matter field are dressed due to the fact that the asymptotic Hamiltonian is not free. The dressing is implemented by conjugation with $e^{R(t)}\ e^{i\Phi(t)}$, where  $\Phi(t)$ is a Hermitian operator that captures the Coulombic force between charged particles and $R(t)$ is an anti-Hermitian operator that describes the ``photon cloud'' surrounding the charged particles. The two operators  commute and  so one can consider their effect separately. It is found that their action on the particle annihilation  operator $b(p)$ produces  a phase:
\ba
e^{-i\Phi(t)} b(p) e^{i\Phi(t)} &  = &  e^{i \alpha_1(p)} b(p) \\
e^{-R(t)} b(p) e^{R(t)} &=& e^{i \alpha_2(p)} b(p) ,
\ea
with\footnote{We adapt the formulas of \cite{KF,akhouryKF} to the normalization of Fock operators used here: $b(p)_{\text{here}}= \sqrt{2 p^0 }(2 \pi)^{3/2}b(p)_{\text{there}}$.}
\ba
\alpha_1(p) & = &  - \ln t  \frac{e^2}{4 \pi}  \int \widetilde{d p'} \frac{p \cdot p'}{\sqrt{(p \cdot p')^2-m^4}} \left(b^\dagger(p') b(p') - c^\dagger(p') c(p') \right) \label{alpha1} \\
\alpha_2(p) & = &  e \int^t d t'    \frac{p^\mu}{p^0} \A_\mu (x^\nu= \frac{t' p^\nu}{p^0}), \label{alpha2}
\ea
where $\widetilde{d p} =\frac{d^3 \vec{p}}{2(2\pi)^3 |\vec{p}|}$  and $\A_\mu$ the free photon field,
\be \label{photonfield}
\A_\mu(x) =  \int \widetilde{d p'}   \,  a_\mu(p') e^{i p' \cdot x} + h.c. 
\ee

The late time asymptotic Kulish-Faddeev  field is then obtained by incorporating the phases in the mode  expansion, which modulo operator ordering is given by\footnote{The phase for $c^\dagger(p)$ should actually appear multiplying on the right. We are ignoring operator orderings since our aim is to compare with the classical expression (\ref{dressedphi}).}
\be \label{phikf}
\varphi_\kf(x) = \int \widetilde{d p}\, e^{i \alpha_1(p) } e^{ i \alpha_2(p)}\left( b(p) e^{i p \cdot x} + c^\dagger(p) e^{-i p \cdot x} \right).
\ee
 To compare with the time-infinity asymptotics used in the paper,  we need to take the  $\tau \to \infty, Y^\mu=$const. limit of (\ref{phikf}). In this limit the momentum integral localizes in $p=m Y$ and one finds
\be \label{phikfti}
\varphi_\kf(\tau,y) \stackrel{\t \to \infty}{=}  e^{i \alpha_1(y) } e^{ i \alpha_2(y)} \varphi_\free(\tau,y)
\ee
where $\alpha(y) \equiv \alpha(p^\mu = m Y^\mu)$.

Comparison with (\ref{dressedphi}) shows that, in order for the two expressions to coincide we should have
\be \label{matchingKF}
\alpha_1(y) +\alpha_2(y) \stackrel{\t \to \infty}{=} e \ln \t A_\t(y).
\ee
We show below that (\ref{matchingKF}) is  satisfied modulo y-independent terms. This suffices for our purposes since the charges depend on $A_\t$ through $\partial_\alpha A_\t$ and is thus insensitive to such ``constant" (constant on the hyperboloid) terms. \footnote{The field strength is also independent of such terms.}

Writing $\alpha_1$ in (\ref{alpha1}) in terms of $p=mY$ and $p'=mY'$, using $\frac{d^3 \vec{p}}{|\vec{p}|}= m^2 d^3 y$ and comparing with (\ref{grey}) one finds
\be \label{alpha10}
\alpha_1(y) = e \ln \t A^j_\t(y) +\frac{e}{4\pi} \ln \t \, \hat{Q} 
\ee
where $\hat{Q}$ is the total electric charge operator. This term corresponds to the  Coulumbic potential due to the total charge, $\A_\t= \frac{1}{\t} \frac{Q}{4 \pi}$. 

We now discuss $\alpha_2$.  Substituting (\ref{photonfield}) in (\ref{alpha2}) and using the prescription $\int^{t}\ d t' e^{is t'}\ =\ \frac{1}{is}e^{is t}$ we can see the $t'$ integral leads to
\be
\alpha_2 =  -i e  \int \widetilde{d p'}\,  \frac{p^\mu}{p \cdot p'} a_\mu(p') e^{i t \frac{p \cdot p'}{p^0}} + h.c.
\ee
We now split the $p'$ integral into angular and energy integrals. Writing $p'^\mu = \w q^\mu$ with $q=(1,\qh)$ the expression becomes
\be
\alpha_2 = -\frac{i e}{2 (2 \pi)^3}  \int d^2 \qh \frac{p^\mu}{p \cdot q} \int_0^\infty d \w a_\mu(\w q) e^{i \w  \frac{p \cdot q}{p^0} t} + h.c.
\ee
Writing 
\be
a_\mu( \w q) \stackrel{\w \to 0^+}{=} \frac{1}{\w} a^{\soft +}_\mu(\qh) + \cdots,
\ee 
the $\w$ integral can be evaluated using $\int_0^\infty d \w \w^{-1}e^{i s \w}=- \ln s$, leading to
\be \label{alpha2b}
\alpha_2 = \ln t \frac{i e }{2 (2 \pi)^3}  \int d^2 \qh \frac{p^\mu }{p \cdot q} a^{\soft +}_\mu(\qh) + h.c.
\ee
To compare with   $\A^\hom_\t$ we need to find the relation between $a_\mu(p')$ and the free data at null infinity. Recall $a_\mu(p')$ in (\ref{photonfield}) is given by
\be
a_\mu(p')= \e^{-}_\mu(p') a_+(p') + \e^{+}_\mu(p') a_-(p')
\ee
where $\e^{\pm}_\mu(p')$ are  positive/negative helicity polarization vectors and $a_\pm(p')$ the corresponding Fock operators. Let us for concreteness consider the polarization vectors used in \cite{strominger}
\be
\e^+_\mu = \partial_z ( \Omega q_\mu), \quad \quad  \e^-_\mu = \partial_{\zb} (\Omega q_\mu)
\ee
where we recall that $\Omega= (1+ z \zb)/\sqrt{2}$   and $p'^\mu = \w q^\mu$ as before. Using the relation between Fock operators and (Fourier transformed) radiative data $A_z$ given in Eq.(\ref{fourier1d}) one finds
\be
a_\mu(\w q) = 4 \pi i \, \Omega^{-1}\ D^A( \Omega q_\mu) \tilde{A}_A(\w,\qh),
\ee
from where  it immediately follows that
\be
a^{\soft +}_\mu(\qh) = - 4 \pi \, \Omega^{-1}\ D^A(\Omega q_\mu) [A_A(\qh)]^+.
\ee
with $[A_A(\qh)]^+$ defined in Eq. (\ref{defsqbA}). Substituting in (\ref{alpha2b}), integrating by parts and evaluating  at $p^\mu = m Y^\mu$ we arrive at
\be
\alpha_2(y) =  e \ln \t \frac{i}{(2 \pi)^2}  \int d^2 \qh \ln(- Y \cdot q) D^A [A_A(\qh)]^+ + h.c.
\ee
Comparing with (\ref{Ahom2}) we conclude that 
\be \label{alpha22}
\alpha_2(y) =  e \ln \t A^j_\t(y) + \textrm{y-independent term}
\ee
where the ``$y$-independent" piece corresponds to the one appearing in Eq. (\ref{feb2-4}).

Thus, Eqs.   (\ref{alpha1}) and (\ref{alpha22}) imply relation (\ref{matchingKF}) holds modulo pure gauge (y-independent) terms.

\section{Conclusion and Open questions}

Analyzing Physics of gauge theories at spatial infinity appears to offer an interesting perspective for analysing the infinite dimensional Asymptotic symmetries which constrain the scattering amplitudes of the theory in the IR sector. Most of the work in this subject has focused on understanding the relationship between soft theorems, Asymptotic symmetries and classical effects like the memory by focusing on symmetries of tree-level scattering amplitudes. For leading soft theorems and associated asymptotic symmetries (like large U(1) gauge transformations at celestial sphere) this is not an issue as soft theorems and the corresponding Asymptotic charges are protected against (infra-red divergent) loop corrections.\footnote{This is no longer true in Non-Abelian gauge theories in four dimensions \cite{bern-loop}} As \cite{sahoo} have shown, the sub-leading soft theorems in Four dimensions gets ``non-perturbatively" corrected by infra-red effects and in fact a new soft factor appears at $O(\ln \omega)$ . In this paper we  have shown that the  story established at tree level, namely an equivalence between infinity of conservation laws and sub-leading soft photon theorem continues to hold even when  sub-leading soft photon theorem receives log corrections. We believe that the charges derived here may be associated to divergent large gauge transformations \cite{subqed,stromqed,prahar} but a detailed analysis of this issue remains to be done. 

Many questions remain open. Although the conserved charges are well defined in classical as well as Quantum theory, their action on radiative data is not clear.  A naive analysis shows that the action of $Q^{\textrm{soft}}[V]$ on the Ashtekar-Struebel radiative data is trivial! Furthermore, the action of $Q^{\textrm{hard}}[V]$ on massive matter does not shed  light on the nature of the symmetry. It is also important to notice that  as the hard charge is quartic in the (matter) field, their Poisson brackets do not appear to close to form an Algebra. Hence the nature of the symmetry associated to the conservation laws derived in this paper remains unclear. 

In this paper we focussed on sub-leading soft photon theorem in the absence of gravitational interaction. As was shown in \cite{sahoo}, the loop corrected soft photon theorem receives further correction once Gravitational drag effect on outgoing soft radiation is taken into account. It will be interesting and highly non-trivial  to understand this modification from the perspective of Asymptotic charges.  This issue is currently under investigation \cite{ssa}.

\section{Acknowledgements}
We are indebted to Nabamita Banerjee, Sayali Atul Bhatkar, Marc Geiler, Arnab Priya Saha, Biswajit Sahoo, Ashoke Sen and Amitabh Virmani for many stimulating discussions. MC would like to thank Chennai Mathematical Institute, Institute of Mathematical Sciences and Raman Research Institute for hospitality during the final stages of this project. 
AL would like to thank Center for High energy Physics (CHEP) in IISC,  Bangalore  for hospitality where part of this work was done. 
AL would like to thank participants of the IISER Pune workshop on soft physics, in particular Shamik Banerjee and Dileep Jatkar for insightful comments. He would also like to thank organisers of the workshop for allowing him to present this work.  
\appendix

\section{Green's functions} \label{greenapp}

One way to obtain the Green's function $\G^\alpha_\beta$ is by expanding in $1/\t$ the standard Minkowski spacetime Green's function.
We start with Maxwell equations in Lorentz gauge,
\be
\square A_\mu = -j_\mu \label{waveeqst} ,
\ee
 and write the corresponding  retarded  solution,
\be
A_\mu(x) = \frac{1}{2 \pi }\int d^4 x' \theta(t-t) \delta((x-x')^2) j_\mu(x'). \label{spacetimegreen}
\ee 
We next express (\ref{spacetimegreen}) in terms of hyperbolic coordinates. Writing $x^\mu = \t Y^\mu$, $x'^\mu= \t' Y'^\mu$  and defining
\be
\sigma:= -Y^\mu Y'_\mu
\ee
one has
\be
d^4 x' = d \t' d^3 y' \t'^3,
\ee
\be
\delta((x-x')^2) = \frac{1}{2 \t \sqrt{\sigma^2-1}}(\delta(\t'-\t_+)+\delta(\t'-\t_-) ) \label{delta}
\ee
where
\be
\t_\pm = \t (\sigma \pm \sqrt{\sigma^2-1}).
\ee
One can check that the $\t_+$ and $\t_-$ roots correspond to advanced or retarded boundary conditions respectively, so only the second one will be relevant in the present discussion.  

The idea now is to substitute the $1/\t'$ expansion of the current on the RHS of  (\ref{spacetimegreen}) and evaluate the  $\t'$ integral with the Dirac delta function to obtain a $1/\t$ expansion for the vector potential. For the $O(e)$ fall-offs, the cartesian components of the current and vector potential have an expansion of the form
\ba
A_\mu(\t,y) & = & \frac{1}{\t} \overset{0}{A}_\mu(y) + \frac{1}{\t^2} \overset{1}{A}_\mu(y) + \cdots  \\
j_\mu(\t,y) & = & \frac{1}{\t^3} \overset{0}{j}_\mu(y) + \frac{1}{\t^4} \overset{1}{j}_\mu(y) + \cdots
\ea
Performing the $\t'$ integral one finds
\be
\overset{1}{A}_\mu(y) = \int d^3 y' g(-Y\cdot Y')\overset{1}{j}_\mu(y') \label{cartesianAj}
\ee
where
\be
g(\sigma) := \frac{1}{4 \pi} \frac{1}{\sqrt{\sigma^2-1}} \frac{1}{(\sigma - \sqrt{\sigma^2-1})}
\ee
From (\ref{cartesianAj}) we can obtain the Green's function $\G_\alpha^\beta$ of Eq. (\ref{FnJn}) as follows. 
The cartesian current components of the current are expressed in terms of hyperbolic components according to
\be
j_\mu= -Y_\mu j_\t+\frac{1}{\t} \D^\alpha Y_\mu j_\alpha,
\ee
whereas the hyperbolic components of the vector potential are
\be
A_\t = Y^\mu A_\mu, \quad A_\alpha = \t \partial_\alpha Y^\mu A_\mu.
\ee
Substituting in (\ref{cartesianAj}) and computing the field strength one eventually arrives at
\be
 \overset{1}{F}_{\t \alpha}(y) =  \int d^3 y' g(\sigma) \D_\alpha \D'^\beta \sigma \overset{0}{j}_\beta(y') - \partial_\alpha  \int d^3 y' g(\sigma) \sigma \overset{1}{j}_\t(y') .\label{Fnjn2}
\ee
This is not explicitly in the form (\ref{FnJn}) but it can be brought into this form by using Eq. (\ref{jtaudiv}) to express  $\overset{1}{j}_\t$ in terms of the divergence  of $\overset{0}{j}_\beta$ and integrating by parts. 

Studying the $\rho \to \infty$ limit of (\ref{Fnjn2}) one finds the desired asymptotic behavior (\ref{Gnbdy}) with
\be
G_B^\alpha(y; \qh) = \frac{1}{8 \pi}  (-q \cdot Y)^{- 3} J^\alpha_{\mu \nu}(y) L^{\mu \nu}_B(\qh) \label{Gbdybulk}
\ee
where $J^\alpha_{\mu \nu} = Y_\mu \D^\alpha Y_\nu - (\mu \leftrightarrow \nu)$ and  $L^{\mu \nu}_B = q^\mu D_B q^\nu - (\mu \leftrightarrow \nu)$ are the Lorentz generators on the hyperboloid and sphere respectively.

Finally, it is interesting  to compute the $\rho \to \infty$ asymptotics of (\ref{Gbdybulk}) by means of the identities \cite{grnfn}
\ba
(-q \cdot Y)^{- 3} & \stackrel{\rho \to \infty}{=}  & 4 \pi \rho  \delta^{(2)}(\xh,\qh) \\
J^A_{\mu \nu}(y) L^{\mu \nu}_B(\qh) |_{\qh = \xh} &\stackrel{\rho \to \infty}{=} & -\rho^{-2} \delta^A_B
\ea
leading to
\be
G_B^A(y; \qh) \stackrel{\rho \to \infty}{=} - \frac{1}{2} \rho^{-1} \delta^A_B  \delta^{(2)}(\xh,\qh).\label{Gbdybdy}
\ee
This limiting behavior can be used to obtain the massless subleading soft factor from the massive one (see comment in footnote \ref{fnmless}).

\subsection{Log-corrected fields}
Repeating the calculation above for the logarithmic corrected $\t$-fall offs one finds
\be
\overset{\ln }{F}_{\t \alpha}(y) = \int d^3 y' \G^\beta_\alpha(y;y') \overset{\ln}{j}_\beta(y') 
\ee
\be
\overset{1}{F}_{\t \alpha}(y) = \partial_\alpha Y^\mu \int d^3 y' (1-\ln(\sigma - \sqrt{\sigma^2-1}))g(\sigma) \overset{\ln}{j}_\mu(y')   + \int d^3 y' \G^\beta_\alpha(y;y') \overset{0}{j}_\beta(y') 
\ee
with
\ba
\overset{\ln }{j}_\mu & = & -Y_\mu \overset{\ln }{j}_\t+ \D^\alpha Y_\mu \overset{\ln }{j}_\alpha \\
&= & D^\alpha (Y_\mu \overset{\ln }{j}_\alpha ) \label{jlndiv},
\ea
where in the last equality we used Eq. (\ref{divjln}). 

Since $\overset{\ln }{F}_{\t \alpha}$ has the same form as the tree-level $\overset{1 }{F}_{\t \alpha}$, with $\overset{\ln}{j}$ playing the role of $\overset{0}{j}_\beta$, we conclude
\be
\overset{\ln }{F}_{\t A}(\rho,\xh) \stackrel{\rho \to \infty}{=} \frac{1}{\rho} \overset{\ln, 0 }{F}_{\t A}(\xh)
\ee
with
\be
\overset{\ln, 0 }{F}_{\t A}(\xh) =  \int d^3 y'   G_A^\beta(y'; \xh)  \overset{\ln}{j}_\beta(y').
\ee
In order to obtain $\overset{0 \ln}{F^+}_{rA}$ from the $(r,u) \leftrightarrow (\t,\rho)$ change of coordinates  we would also need to discuss the $\rho \to \infty$ expansion of $\overset{1}{F}_{\t A}$. Since this procedure is quite cumbersome,  we  follow in the next section  a simpler route for obtaining   $\overset{0 \ln}{F^+}_{rA}$ in terms of the current at time-infinity.
\subsection{Eq. (\ref{feb1-1})
} \label{Frzapp}
From the field equation at null infinity (\ref{FrzAz}) we know the $\ln u$ coefficient of $\overset{0}{F}_{rz}(u,x)$ is related with the $1/u$ coefficient of $A_z(u,\xh)$ by
\be \label{FrzAzplus}
\overset{0 \ln}{F}_{rz}(\xh) = - D_z D^{\zb} \overset{1}{A}_{\zb}(\xh)
\ee
The idea now is  to evaluate $\overset{1}{A}_{\zb}$ directly from (\ref{spacetimegreen}). For this purpose we  write $x^\mu$ in retarded coordinates and $x'^\mu$ in hyperboloid coordinates:
\ba
x^\mu & =  & r (1 , \xh) + u(1,\vec{0}) \\
x'^\mu & = & \t Y^\mu
\ea
introducing the notation $q^\mu = (1, \xh)$ we have 
\be
-(x-x')^2 = 2 r(u + \t' q \cdot Y' ) + O(r^{0}).
\ee
Substituting in (\ref{spacetimegreen}) we find 
\be
A_\mu(x)  =  \frac{1}{4 \pi r} \int d \t d^3 y \t^3 \delta(u+ \t q \cdot Y)  j_\mu(\t,y) + O(1/r^2)
\ee
We now consider the  $\t \to \infty$ expansion of the current
\be
\t^3 j_\mu(\t,y) = \overset{0}{j}_\mu(y) + \frac{\ln \t}{\t}  \overset{\ln}{j}_\mu(y) + \frac{1}{\t}  \overset{1}{j}_\mu(y) + \cdots
\ee
and evaluate the $\t$ integral with the delta function. This yields  a $u \to \infty$ expansion  of the form
\be
A_\mu(x) = \frac{1}{r} \left( \overset{0}{A}_\mu(\xh)+ \frac{\ln u}{u} \overset{\ln}{A}_\mu(\xh) + \frac{1}{u} \overset{1}{A}_\mu(\xh) + \cdots \right) + O(1/r^2)
\ee
There appears to be a new $\ln u /u$ term, however
\be \label{Alnru}
\overset{\ln}{A}_\mu(\xh)  = \frac{1}{4 \pi} \int d^3 y \overset{\ln}{j}_\mu(y)  =0
\ee
because $\overset{\ln}{j}_\mu$ is a total derivative (Eq. (\ref{jlndiv})). 
The term of interest for us is 
\be 
 \overset{1}{A}_\mu(\xh) = \frac{1}{4 \pi} \int d^3 y \left(   -  \ln(- q \cdot Y) \overset{\ln}{j}_\mu(y) + \overset{1}{j}_\mu(y)  \right), \label{A1}
\ee
with $\overset{\ln}{j}_\mu$ given in (\ref{jlndiv}) and 
\ba
 \overset{1}{j}_\mu & = & - Y_\mu \overset{1}{j}_\t + \D^\alpha Y_\mu \overset{0}{j}_\alpha \\
&=& Y_\mu \D^\alpha \overset{\ln}{j}_\alpha + \D^\alpha(Y_\mu \overset{0}{j}_\alpha)
\ea
where in the second equality we used Eq. (\ref{divjln}) and $\overset{1}{j}_\t = \overset{\ln}{j}_\t -\D^\alpha \overset{0}{j}_\alpha $ as implied by current conservation. 
Substituting these expressions in (\ref{A1}) and rearranging  terms one finds $\overset{1}{A}_\mu$ can be written as
\be
\overset{1}{A}_\mu(\xh) =  \frac{1}{4 \pi} \int d^3 y \, (q \cdot Y)^{-1} q^\nu J^\alpha_{\mu \nu}(y) \,  \overset{\ln}{j}_\alpha(y)
\ee
where $J^\alpha_{\mu \nu}(y)= Y_\mu \D^\alpha Y_\nu - Y_\nu \D^\alpha Y_\mu$.
We finally need to evaluate 
\be
\overset{1}{A}_{\zb} = \partial_{\zb} q^\mu \overset{1}{A}_\mu
\ee
Comparing with (\ref{S1}) we see that $\overset{1}{A}_{\zb}$ is essentially given by the tree-level subleading soft factor. We can then use the identity (\ref{keyid}) to evaluate the sphere derivatives (\ref{FrzAzplus}) (written as $D_z D^{\zb}= \gamma_{z \zb } D^2_z$). This leads to 
\be
\overset{0 \ln}{F}_{rz}(\xh) =    - \frac{1}{8 \pi} \int d^3 y \, (q \cdot Y)^{-3}  L^{\mu \nu}_z(\xh) J^\alpha_{\mu \nu}(y)  \overset{\ln}{j}_\alpha(y).
\ee


\end{document}